\documentclass[11pt,a4paper]{article}

\usepackage[a4paper,margin=2.5cm]{geometry}
\usepackage[utf8]{inputenc}
\usepackage{bbm,braket,amsmath,amsfonts,amsthm,float,tikz,thm-restate,tabularx,booktabs,quantikz,url,amssymb,multirow}
\usepackage[english]{babel}
\usepackage[hypertexnames=false,colorlinks=true,linkcolor=magenta,urlcolor=blue,citecolor=blue]{hyperref}
\usepackage[capitalize,nameinlink]{cleveref}
\usepackage{authblk}
\usepackage{pgfplots,pgfplotstable}
\pgfplotsset{compat=1.18}
\definecolor{forestgreen}{rgb}{0.13, 0.55, 0.13}

\usepackage{algorithm}
\usepackage[noEnd=true,indLines=true]{algpseudocodex}
\algnewcommand{\Input}{\item[\textbf{Input:}]}
\algrenewcommand{\Output}{\item[\textbf{Output:}]}

\usepackage{thmtools}
\declaretheorem[style=plain,numberwithin=section]{theorem}
\declaretheorem[style=plain,numberlike=theorem]{lemma,corollary}

\declaretheorem[style=definition,numberlike=theorem]{definition}

\declaretheorem[style=plain,numberlike=theorem]{fact}

\declaretheorem[style=plain,numberlike=theorem]{observation}

\Crefname{lemma}{Lemma}{Lemmas}
\Crefname{fact}{Fact}{Facts}
\Crefname{theorem}{Theorem}{Theorems}
\Crefname{corollary}{Corollary}{Corollaries}
\Crefname{claim}{Claim}{Claims}
\Crefname{example}{Example}{Examples}
\Crefname{problem}{Problem}{Problems}
\Crefname{definition}{Definition}{Definitions}
\Crefname{notation}{Notation}{Notations}
\Crefname{assumption}{Assumption}{Assumptions}
\Crefname{subsection}{Section}{Sections}
\Crefname{observation}{Observation}{Observations}
\Crefname{section}{Section}{Sections}
\Crefformat{equation}{(#2#1#3)}

\newcommand{\DQC}{\mathsf{DQC}}
\newcommand{\BPP}{\mathsf{BPP}}

\newcommand{\PP}{\mathsf{PP}}
\newcommand{\GapP}{\mathsf{GapP}}

\DeclareMathOperator{\polylog}{polylog}
\DeclareMathOperator{\poly}{poly}

\DeclareMathOperator{\tr}{tr}

\newcommand{\N}{\mathbbm N}

\newcommand{\C}{\mathbbm C}

\begin{document}

\title{Dequantization and Hardness of Spectral Sum Estimation}

\author[1]{Roman Edenhofer\thanks{edenhofer@irif.fr}}
\author[2]{Atsuya Hasegawa\thanks{atsuya.hasegawa@math.nagoya-u.ac.jp}}
\author[2]{François Le Gall\thanks{legall@math.nagoya-u.ac.jp}}
\affil[1]{Université Paris Cité, CNRS, IRIF, Paris, France}
\affil[2]{Graduate School of Mathematics, Nagoya University, Japan}

\date{}

\maketitle


\begin{abstract}
We give new dequantization and hardness results for estimating spectral sums of matrices, such as the log-determinant. Recent quantum algorithms have demonstrated that the logarithm of the determinant of sparse, well-conditioned, positive matrices can be approximated to $\varepsilon$-relative accuracy in time polylogarithmic in the dimension $N$, specifically in time $\mathrm{poly}(\mathrm{log}(N), s, \kappa, 1/\varepsilon)$, where $s$ is the sparsity and $\kappa$ the condition number of the input matrix. We provide a simple dequantization of these techniques that preserves the polylogarithmic dependence on the dimension. Our classical algorithm for the log-determinant runs in time~\mbox{$\mathrm{polylog}(N)\cdot s^{O(\sqrt{\kappa}\log(\kappa/\varepsilon))}$} which constitutes an exponential improvement over previous classical algorithms in certain parameter regimes.

We complement our classical upper bounds with complexity-theoretic limitations. We prove that estimating normalized traces of polynomial powers and inverses of log-local Hamiltonians to inverse-polynomial additive accuracy is $\DQC1$-complete, resolving an open problem of Cade and Montanaro (TQC 2018) concerning the complexity of Schatten-$p$ norm estimation. Finally, we prove a general $\PP$-completeness result for unnormalized spectral sums: under mild polynomial-approximability and nondegeneracy assumptions on $f$, estimating $\tr[f(A)]$ to constant additive accuracy is $\PP$-complete.
\end{abstract}

\clearpage
\tableofcontents
\clearpage

\section{Introduction}

\subsection{Background}

For a Hermitian matrix $A\in\mathbb{C}^{N\times N}$ and a function $f$ defined on the spectrum of $A$, the spectral sum is
\[
    \tr[f(A)] = \sum_{i=1}^N f(\lambda_i),
\]
where $\lambda_1,\dots,\lambda_N$ are the eigenvalues of $A$.
A central example, and the main focus of this work, is the log-determinant
\[
    \log\det(A) = \log\Bigl(\prod_{i=1}^N \lambda_i\Bigr) = \sum_{i=1}^N \log \lambda_i.
\]
An $\varepsilon$-additive approximation to $\log\det(A)$ corresponds to an $(e^{\varepsilon}-1)\approx \varepsilon$ multiplicative approximation of the determinant of $A$, for small $\varepsilon>0$.
Other spectral sums of interest include the partition function $Z(A,\beta) = \tr[e^{-\beta A}]$ with inverse temperature $\beta>0$, the trace of powers~$\tr[A^p]$, which corresponds to the $p$-th power of Schatten $p$-norms for positive matrices, and the trace of the inverse $\tr[A^{-1}]$.

Spectral sums arise in diverse areas such as machine learning, computational chemistry, biology, statistics, and finance (see, e.g., \cite{ABB00,WR06,NHD12}), and many classical estimation algorithms have been proposed (see, e.g., \cite{HMS15,UCS17,MMMW21}).
Most of these methods build on Hutchinson’s stochastic trace estimator \cite{Hutch89}: for a random vector $v\in\mathbb{R}^N$ with~$\mathbb{E}[vv^T]=I$, one has~$\tr[A] = \mathbb{E}[v^T A v]$.
In practice,~$v$ is often drawn from the Rademacher distribution which minimizes the estimators variance. The time to evaluate~$v^T A v$ is then determined by the time for matrix–vector multiplication and scales linearly in the number of nonzero entries of~$A$.
For instance, the fastest known classical algorithm estimates the log-determinant with relative error~$\varepsilon$ in time
\[
    O\!\left(\frac{\sqrt{\kappa}\,\|A\|_0}{\varepsilon^2} \log\Bigl(\frac{\kappa}{\varepsilon\delta}\Bigr)\right),
\]
where~$\delta$ is the failure probability and $\|A\|_0$ the number of nonzeros of $A$ \cite{HMS15}.
Even if $A$ is sparse with at most $s$ nonzeros per row and column, $\|A\|_0$ typically scales polynomially with $N$.

Recently, several quantum algorithms for estimating spectral sums, most notably the log-determinant, have been proposed \cite{ZFO+19,LS24,GLM25}.
For sparse matrices, these algorithms avoid the polynomial dependence on the matrix dimension $N$, achieving instead an~$\varepsilon$-relative approximation of~$\log\det(A)$ in time
\[
    \poly(\log N,s,\kappa,1/\varepsilon),
\]
where~$s$ is the sparsity and~$\kappa$ the condition number of the input matrix.
The algorithms of~\cite{ZFO+19,GLM25} use Hamiltonian simulation and phase estimation to sample eigenvalues from the maximally mixed state. The approach of \cite{LS24} instead uses the quantum singular value transformation \cite{GSLW19} to implement a block-encoding of a polynomial approximation of $\log(A)$, whose trace is then estimated via amplitude estimation applied to a Hadamard-test circuit.

These algorithms are closely related to the model of quantum computation with one clean qubit, known as $\DQC1$, introduced by Knill and Laflamme \cite{KL98} in the context of high-temperature NMR (nuclear magnetic resonance) quantum devices.
In this model, one applies a polynomial-time quantum circuit to the $n$-qubit state
\[
    \rho = \ket{0}\bra{0} \otimes \tfrac{1}{2^{n-1}} I_{n-1},
\]
where a single qubit is initialized in $\ket{0}$ and the remaining qubits are maximally mixed.
Cade and Montanaro \cite{CM17} showed that normalized spectral sums of sufficiently regular functions can be estimated in $\DQC1$ for log-local Hamiltonians. Their framework includes the log-determinant, partition function, traces of powers, and trace of the inverse under corresponding spectral promises. Brand{\~a}o \cite{Bra08} proved matching $\DQC1$-hardness of estimating the normalized partition function at polynomial inverse temperature.

These developments raise two natural questions. First, to what extent can the runtime parameters of the quantum algorithms be reproduced classically? This question was posed explicitly by Aaronson \cite{Aar25} in response to the log-determinant algorithm of \cite{GLM25}. Second, when such a dequantization is not possible, can one identify a complexity-theoretic obstruction?

Our work addresses both questions. It also fits into the broader program of dequantizing quantum linear-algebra algorithms, initiated by Tang \cite{Tan19} and subsequently extended to several algorithms based on the quantum singular value transformation \cite{CGL+22,GLG23,BT24,LG25}.

\subsection{Our results}
We write $N=2^n$ and consider matrices specified through \emph{sparse access}: given a row or column index, one can enumerate its at most $s$ nonzero entries in time $\poly(n)$. We write $O^*(T)$ to suppress such factors polynomial in $n$, as well as polylogarithmic overheads.

We give two classical algorithmic frameworks: one for sparse matrices and one for local Hamiltonians. Our first framework is summarized by the following informal theorem.

\begin{theorem}[Informal version of \cref{thm: deterministic powering trace of general function}] 
    Let $f:I\to\mathbb{R}$ admit an efficiently computable degree~$d$ polynomial approximation with uniform error $\varepsilon$ on $I$. 
    Then, for every $s$-sparse Hermitian matrix $A\in\mathbb{C}^{2^n\times2^n}$ with spectrum in $I$, one can estimate, with high probability,
    \[
        \frac{1}{2^n}\tr[f(A)]\pm2\varepsilon
            \qquad\text{in time}\qquad O^*\!\left(s^d f_{\max}^2/\varepsilon^2\right),
    \]
    where $f_{\max}:=\sup_{x\in I}|f(x)|$.
\end{theorem}

The theorem shows that the central analytic parameter to estimate normalized spectral sums of sparse matrices is the degree required to approximate~$f$ on the spectrum of $A$. In particular, a constant-degree approximation immediately gives a classical algorithm with running time polynomial in $s$ and polylogarithmic in $N$.

For the logarithm on $[1/\kappa,1]$, a Chebyshev approximation of degree
\[
    d=O\!\left(\sqrt{\kappa}\log(\kappa/\varepsilon)\right)
\]
gives an $(\varepsilon/\kappa)$-additive approximation of the normalized log-determinant. 
Under a standard spectral upper bound away from $1$, this can be converted into an $\varepsilon$-relative approximation of the unnormalized log-determinant.

\begin{corollary}[Informal version of \cref{cor: relative error classical logdet approx}]
    Let $A\in\mathbb{C}^{2^n\times 2^n}$ be an $s$-sparse Hermitian matrix with spectrum in $[\frac{1}{\kappa},1-\frac{1}{\kappa}]$.
    One can compute, with high probability, a relative estimate
    \[
        \log\det(A) \pm\varepsilon|\log\det(A)|
            \qquad \text{ in time} \qquad
            O^*(s^{c\cdot\sqrt{\kappa}\log(\kappa/\varepsilon)}),
    \]
    where $c>1$ is a universal constant.
\end{corollary}

Thus, when $s=\poly(n)$ and $\kappa$ and $\varepsilon$ are sufficiently small, the running time is polynomial in $n=\log N$, although the matrix may contain $N\polylog(N)$ nonzero entries. In these regimes, our algorithm gives an exponential improvement in the dimension over methods whose running time scales with $\|A\|_0$, such as \cite{HMS15}. The result only partially dequantizes the quantum algorithms, however, as our runtime is not polynomial in all the parameters $s,\kappa$ and~$\varepsilon^{-1}$.

Our second algorithmic framework exploits additional structure when the input is a local Hamiltonian. Let $H=\sum_{i=1}^{m}H_i$ be $k$-local and normalized so that its total interaction strength is bounded
\[
    \sum_{i=1}^m\|H_i\|_2\leq1.
\]
Although $H$ may have as many as $m\cdot2^k$ nonzero entries per row, its induced matrix $1$-norm can be bounded in terms of $k$ alone. A random-walk estimator then removes the dependence on the number of local terms $m$ from the exponential part of the running time.

\begin{theorem}[Informal version of \cref{thm: random walk trace of function for Hamiltonians}]
    Let $f:I\to\mathbb{R}$ admit a degree~$d$ polynomial approximation $p(x)=\sum_{k=0}^d c_k x^k$ with uniform error $\varepsilon$ on $I$. 
    For every $k$-local Hamiltonian with spectrum in $I$ and bounded total interaction strength, one can estimate, with high probability, 
    \[ 
        \frac{1}{2^n}\tr[f(H)]\pm2\varepsilon 
            \qquad\text{in time}\qquad  O^*\!\left(2^{kd}\|p\|_1^2/\varepsilon^2\right), 
    \]
    where $\|p\|_1 = \sum_{k=0}^d |c_k|$ denotes the coefficient norm of $p$.
\end{theorem}

Combining the two general algorithms with polynomial approximations for
\[
    \log x,\qquad e^{-\beta x},\qquad x^p,
    \qquad\text{and}\qquad x^{-1}
\]
gives classical algorithms for the normalized log-determinant, partition function, trace of powers, and trace of the inverse. We state these applications in \cref{cor: classical applications}.

We next investigate whether substantially better classical algorithms should be expected. For log-local Hamiltonians, Cade and Montanaro \cite{CM17} established $\DQC1$ containment for all four spectral sums above in the parameter regime where $\varepsilon=1/\poly(n)$ and $\kappa,\beta,p=\poly(n)$. 
Before this work, the matching $\DQC1$-hardness has only been shown in the case of the normalized partition function by Brand{\~a}o \cite{Bra08}. We prove matching hardness for traces of polynomial powers and for the trace of the inverse.

\begin{theorem}[cf. \cite{CM17} and \cref{thm: trace A inverse is DQC1-hard,thrm: trace H^p is DQC1-hard}]
    For log-local Hamiltonians $H$ acting on~$n$ qubits, estimating \vspace{-.4cm}
    \[
        \begin{aligned} 
            \frac{1}{2^n}\tr[H^p] &\quad\text{when}\quad \sigma(H)\subseteq[-1,1], \qquad\text{or}\\
            \frac{1}{2^n}\tr[H^{-1}] &\quad\text{when}\quad \sigma(H)\subseteq[1/\kappa,1], 
        \end{aligned} 
    \] 
    to additive accuracy $\varepsilon=1/\poly(n)$ is $\DQC1$-complete for $\kappa,p=\poly(n)$. 
\end{theorem}

Consequently, unless $\DQC1\subseteq\BPP$, there is no polynomial-time classical algorithm for these problems throughout the full parameter regime of the $\DQC1$-containment.
Cade and Montanaro previously proved hardness for traces of powers only under a substantially stronger accuracy requirement. Our result matches their containment and thus resolves the most immediate open problem in their paper concerning the complexity of Schatten-$p$ norm estimation. After the first version of this work appeared, Shao and Wang \cite{SW25} established the analogous $\DQC1$-hardness result for the log-determinant, resolving the last missing piece in the complexity of these four spectral sums.
All of these hardness results have been generalized to spectral sums where $f$ satisfies a polynomial lower bound on its approximate degree in \cite{JLSWZ26}.

Finally, we characterize the complexity of estimating unnormalized spectral sums. Our result applies whenever the function satisfies a mild polynomial-approximability and non-degeneracy constraint, which are in particular satisfied for all four spectral sums above.

\begin{theorem}[Informal version of
\cref{thm: PP-containment,thm: PP-hardness}]
    Let $f:I\to\mathbb{R}$ satisfy the polynomial-approximability and non-degeneracy assumptions of \cref{thm: PP-containment,thm: PP-hardness}.
    Then, given a polynomially sparse Hermitian matrix $A\in\mathbb{C}^{2^n\times2^n}$ with spectrum in $I$, estimating the unnormalized spectral sum $\tr[f(A)]$ to constant additive accuracy is $\PP$-complete.
\end{theorem}

Equivalently, the theorem concerns estimating the normalized quantity $2^{-n}\tr[f_n(A)]$ to inverse-exponential additive accuracy $\Theta(2^{-n})$. 
In fact, while the hardness holds for constant additive error for the unnormalized spectral sum, the containment remains valid for promise gaps as small as $2^{-\poly(n)}$.

The resulting complexity landscape for the four normalized spectral sums is summarized in \cref{tab: complexity summary}. At constant spectral parameters and constant additive error, our classical algorithms place all four problems in $\BPP$. For inverse-polynomial accuracy and polylogarithmic condition number, inverse temperature, or power, they yield classical quasipolynomial-time algorithms. By contrast, when these spectral parameters are polynomial, the corresponding problems for log-local Hamiltonians are $\DQC1$-complete.
To our knowledge, no classical quasipolynomial-time algorithm is known for simulating general $\DQC1$ computations. Thus, the apparent quantum advantage seems to not arise from achieving higher estimation accuracy, but rather from these spectral parameters. Finally, at inverse-exponential additive accuracy, the problems become~$\PP$-complete.

\begin{table}[t]
    \centering
    \fontsize{9pt}{10.2pt}\selectfont
    \renewcommand{\arraystretch}{1.5}
    \setlength{\tabcolsep}{2.5pt}
    \begin{tabular}{@{}c|l|c|l@{}}
        \hline
        \begin{tabular}{@{}c@{}}
            \textbf{spectral sum}\\
            (\textbf{spectral promise})
        \end{tabular}
        &
        \multicolumn{1}{c|}{\textbf{parameters}}
        &
        \textbf{complexity}
        &
        \multicolumn{1}{c}{\textbf{references}}
        \\
        \hline\hline

        \begin{tabular}{@{}c@{}}
            $\frac{1}{2^n}\log\det(A)\pm\varepsilon$\\
            ($\sigma(A)\subseteq [\frac1\kappa,1]$)
        \end{tabular}
        
        &
        \begin{tabular}{@{}l@{}}
            $\kappa,\varepsilon^{-1}=O(1)$\\
            $\kappa=\polylog(n)$, $\varepsilon^{-1}=\poly(n)$\\
            $\kappa=\poly(n)$, $\varepsilon^{-1}=\poly(n)$\\
            $\kappa=\poly(n)$, $\varepsilon^{-1}=2^{\poly(n)}$
        \end{tabular}
        &
        \begin{tabular}{@{}c@{}}
            $\in\BPP$\\
            classical quasi-polytime\\
            $\DQC1$-complete\\
            $\PP$-complete
        \end{tabular}
        &
        \begin{tabular}{@{}l@{}}
            \cref{cor: classical applications}\\
            \cref{cor: classical applications}\\
            \cite{CM17,SW25}\\
            \cref{lem: log-approx},\;\cref{thm: PP-containment,thm: PP-hardness}
        \end{tabular}
        \\
        \hline

        \begin{tabular}{@{}c@{}}
            $\frac{1}{2^n}\tr[e^{-\beta A}]\pm\varepsilon$\\
            ($\sigma(A)\subseteq [0,1]$)
        \end{tabular}
        &
        \begin{tabular}{@{}l@{}}
            $\beta,\varepsilon^{-1}=O(1)$\\
            $\beta=\polylog(n)$, $\varepsilon^{-1}=\poly(n)$\\
            $\beta=\poly(n)$, $\varepsilon^{-1}=\poly(n)$\\
            $\beta=\poly(n)$, $\varepsilon^{-1}=2^{\poly(n)}$
        \end{tabular}
        &
        \begin{tabular}{@{}c@{}}
            $\in\BPP$\\
            classical quasi-polytime\\
            $\DQC1$-complete\\
            $\PP$-complete
        \end{tabular}
        &
        \begin{tabular}{@{}l@{}}
            \cref{cor: classical applications}\\
            \cref{cor: classical applications}\\
            \cite{Bra08,CM17,CSS21}\\
            \cref{lem: exp-approx},\;\cref{thm: PP-containment,thm: PP-hardness}
        \end{tabular}
        \\
        \hline

        \begin{tabular}{@{}c@{}}
            $\frac{1}{2^n}\tr[A^p]\pm\varepsilon$\\
            ($\sigma(A)\subseteq [-1,1]$)
        \end{tabular}
        &
        \begin{tabular}{@{}l@{}}
            $p,\varepsilon^{-1}=O(1)$\\
            $p=\polylog(n)$, $\varepsilon^{-1}=\poly(n)$\\
            $p=\poly(n)$, $\varepsilon^{-1}=\poly(n)$\\
            $p=\poly(n)$, $\varepsilon^{-1}=2^{\poly(n)}$
        \end{tabular}
        &
        \begin{tabular}{@{}c@{}}
            $\in\BPP$\\
            classical quasi-polytime\\
            $\DQC1$-complete\\
            $\PP$-complete
        \end{tabular}
        &
        \begin{tabular}{@{}l@{}}
            \cite{CM17,AGSS23}, \cref{cor: classical applications}\\
            \cite{CM17,AGSS23}, \cref{cor: classical applications}\\
            \cite{CM17}, \cref{thrm: trace H^p is DQC1-hard}\\
            \cref{lem: monomial-approx},\;\cref{thm: PP-containment,thm: PP-hardness}
        \end{tabular}
        \\
        \hline

        \begin{tabular}{@{}c@{}}
            $\frac{1}{2^n}\tr[A^{-1}]\pm\varepsilon$\\
            ($\sigma(A)\subseteq [\frac1\kappa,1]$)
        \end{tabular}
        &
        \begin{tabular}{@{}l@{}}
            $\kappa,\varepsilon^{-1}=O(1)$\\
            $\kappa=\polylog(n)$, $\varepsilon^{-1}=\poly(n)$\\
            $\kappa=\poly(n)$, $\varepsilon^{-1}=\poly(n)$\\
            $\kappa=\poly(n)$, $\varepsilon^{-1}=2^{\poly(n)}$
        \end{tabular}
        &
        \begin{tabular}{@{}c@{}}
            $\in\BPP$\\
            classical quasi-polytime\\
            $\DQC1$-complete\\
            $\PP$-complete
        \end{tabular}
        &
        \begin{tabular}{@{}l@{}}
            \cref{cor: classical applications}\\
            \cref{cor: classical applications}\\
            \cite{CM17}, \cref{thm: trace A inverse is DQC1-hard}\\
            \cref{lem: inv-approx},\;\cref{thm: PP-containment,thm: PP-hardness}
        \end{tabular}
        \\
        \hline
    \end{tabular}
    \caption{
        Complexity landscape of the spectral sums considered in this work.
        The $\DQC1$ results apply to log-local Hamiltonians, while the $\BPP$ and $\PP$ results apply to polynomially sparse matrices.
    }
    \label{tab: complexity summary}
\end{table}

\newpage
\subsection{Our techniques}

\subsubsection*{Our classical algorithms}

Both classical algorithms begin with the same observation: 
If a degree~$d$ polynomial $p$ uniformly approximates $f$ on the spectrum of $A$, then the corresponding normalized spectral sums of $f$ and $p$ are also close, since
\[
    \left|\frac{1}{2^n}\tr[f(A)] - \frac{1}{2^n}\tr[p(A)]\right|
        \leq \|f-p\|_{\infty}.
\]
It therefore suffices to estimate the normalized trace of $p(A)$.

The key elementary identity is
\[
    \frac{1}{2^n}\tr[p(A)]
        =
        \mathbb{E}_{i\in[2^n]}[p(A)(i,i)].
\]
In both algorithms, we sample a uniformly random index $i$ and estimate the diagonal entry~$p(A)(i,i)$. Since
\[
    p(A)(i,i)=\sum_{k=0}^{d}c_kA^k(i,i),
\]
the task reduces to computing diagonal entries of powers of $A$.

For an $s$-sparse matrix, $A^k(i,i)$ is a sum over closed walks of length $k$ beginning and ending at $i$. Our first algorithm explicitly enumerates these walks. There are at most $s^k$ of them, so all powers up to degree $d$ can be evaluated in time $O^*(s^d)$. Sampling independent diagonal entries and applying a concentration bound yields the correctness and runtime of the first algorithm.

The second algorithm replaces exhaustive enumeration by a random walk. Starting from a uniformly random index $i$, it performs a walk whose transition probabilities are proportional to the absolute values of the matrix entries. A suitable product of row norms and complex signs makes the contribution of each sampled closed walk an unbiased estimator of $A^k(i,i)$. Its magnitude is controlled by the $1$-norm of $A$ and with a concentration bound gives a total runtime of $O^*(\max\{1,\|A\|_1^{2d}\}\|p\|_1^2/\varepsilon^2)$.
The algorithm is essentially identical to the one presented by Apers, Gribling, Sen and Szabó \cite{AGSS23}.
We generalize it to arbitrary spectral sums and observe that it is especially useful for $k$-local Hamiltonians
\[
    H=\sum_{i=1}^m H_i
    \qquad\text{satisfying}\qquad
    \sum_{i=1}^m\|H_i\|_2\leq1.
\]
Each local term satisfies $\|H_i\|_1\leq2^{k/2}\|H_i\|_2$, and hence $\|H\|_1\leq2^{k/2}$.
The resulting running time in that case is therefore $O^*(2^{kd}\|p\|_1^2/\varepsilon^2)$, independent on the number of local terms $m$.

The remaining ingredient for our applications is approximation theory. We use low-degree polynomial approximations via Chebyshev truncation for the logarithm, exponential, reciprocal, and monomials. Substituting their degrees into the two general algorithms yields fast algorithms for the normalized log-determinant, partition function and traces of powers and inverses.

Conceptually, our first algorithm, based on deterministic path enumeration, can also be viewed as an application of the dequantization of the quantum singular value transformation developed in \cite{GLG23}. 
For the maximally entangled state $\ket{\Phi} = \frac{1}{\sqrt{2^n}} \sum_{i\in[2^n]}\ket{i}\ket{i}$, one finds 
\[ 
    \bra{\Phi} \bigl(f(A)\otimes I\bigr)\ket{\Phi} \;=\; \frac{1}{2^n}\sum_{i,j=1}^{2^n}\bra{i}f(A)\ket{j}\braket{i|j} \;=\; \frac{1}{2^n}\sum_{i=1}^{2^n}f(A)(i,i) \;=\; \frac{1}{2^n}\tr[f(A)].
\] 
Theorem~4.1 of \cite{GLG23} can therefore be applied to a polynomial approximation of $f(A)$ to yield an estimate of the normalized spectral sum. This dequantization viewpoint motivated our approach, but in the present setting the resulting classical procedure admits the simpler elementary interpretation described above: sample a diagonal entry uniformly and evaluate it by sparse matrix powering.

\subsubsection*{$\DQC1$-completeness for log-local Hamiltonians}

Our hardness proofs rely on Brand{\~a}o's adaptation \cite{Bra08} of Kitaev's circuit-to-Hamiltonian construction \cite{KSV02}. Given a $\DQC1$ circuit, the construction yields a log-local Hamiltonian whose lowest $2^{n-1}$ eigenvalues are separated from the rest of the spectrum by a controllable gap and whose average encodes the circuit's rejection probability.

The partition function is particularly well suited to this construction because $x\mapsto e^{-\beta x}$ suppresses the eigenvalues above the gap. Brand{\~a}o used this observation to prove $\DQC1$-hardness of normalized partition-function estimation.

Our proofs follow and extend the same template. We first rescale and shift the Hamiltonian so that its spectrum satisfies the promise required by the target problem. We then choose the parameters so that the relevant function has two properties: on the low-energy part of the spectrum, it is well approximated by an affine function whose linear term reveals the average low-lying eigenvalue, and on the remaining spectrum, its total contribution to the normalized trace is negligible.

For the trace of a power, we apply the affine transformation
\[
    \widehat{H}=I-\beta H.
\]
With an appropriate choice of parameters, the function $x\mapsto(1-\beta x)^p$ behaves approximately affinely on the relevant low-energy window while strongly suppressing eigenvalues above the spectral gap. For the trace of the inverse, we use the same general template with a different shift of the Hamiltonian. In both cases, suitable parameter choices yield inverse-polynomial additive hardness matching the known $\DQC1$ upper bounds.

\subsubsection*{$\PP$-completeness}

For the $\PP$ upper bound, we first replace $f$ by a polynomial $p$ whose uniform error is small enough so that
\[
    |\tr[f(A)]-\tr[p(A)]|
\]
is small. Because the trace is unnormalized, this requires approximating $f$ to accuracy roughly~$2^{-n}$ times the desired trace accuracy. Our assumptions guarantee that the resulting polynomial still has polynomial degree.

We then expand $\tr[p(A)]$ as a weighted sum over closed sparse walks. There are exponentially many such walks, but we find that each individual contribution can be computed in polynomial time. We encode the resulting signed sum as a $\GapP$ function. Comparing its sign with an appropriately scaled threshold gives containment in $\PP$.

The hardness reduction is direct. Given a Boolean formula $\varphi$, we construct a one-sparse diagonal matrix $A_{\varphi}$ whose diagonal entry is $\alpha$ on satisfying assignments and $\beta$ on unsatisfying assignments. If $f(\alpha)$ and $f(\beta)$ differ by a constant, then
\[
    \tr[f(A_\varphi)]
\]
is an affine function of the number of satisfying assignments of $\varphi$. A constant-additive estimate therefore distinguishes whether a majority of assignments satisfy $\varphi$, giving a reduction from~$\mathrm{MAJSAT}$.

\subsection{Open questions}
Our $\DQC1$ upper bounds apply to log-local Hamiltonians, whereas the classical and $\PP$ results apply to general polynomially sparse matrices. It remains open whether normalized spectral sums of general sparse matrices can also be estimated in $\DQC1$.

A natural route would be an ancilla-efficient block-encoding or Hamiltonian-simulation procedure for a sparse $2^n\times2^n$ matrix. Standard sparse-matrix constructions \cite{AT03,BCK15,GSLW19} use workspace whose size grows with $n$, while a $\DQC1$ computation has only logarithmically many clean qubits available after standard amplification. Extending the $\DQC1$ algorithms beyond log-local Hamiltonians may therefore require a substantially more economical implementation of sparse-access oracles, or an approach that avoids block-encodings altogether.

A second question is whether the dependence of our classical algorithms on the polynomial-approximation degree can be improved. Our hardness results rule out polynomial-time classical algorithms throughout the full polynomial-parameter regime under the assumption~\mbox{$\DQC1\nsubseteq\BPP$}, but leave open faster classical algorithms in intermediate regimes, for example when $\kappa$, $\beta$, or $p$ grow polylogarithmically with $n$.

\subsection{Organization of this paper}

In \cref{sec:preliminaries}, we introduce notation, the matrix-access models, complexity classes used throughout the paper and collect some probabilistic and analytic preliminaries.

In \cref{sec:classical_algorithm}, we present our two classical algorithms for estimating normalized spectral sums, and derive their applications to the log-determinant, partition function, and traces of powers and inverses.

In \cref{sec:DQC1}, we recall the $\DQC1$ containment result of Cade and Montanaro \cite{CM17} and prove matching hardness for traces of polynomial powers and inverses.

Finally, in \cref{sec: PP-completeness}, we establish the general $\PP$ containment and hardness results for unnormalized spectral sums.

\section{Preliminaries}\label{sec:preliminaries}

We collect the notation, probabilistic and analytic tools, complexity classes,
and matrix-access models used throughout the paper. For general background on
quantum computation, we refer to \cite{NC10}.

\subsection{Notation and norms}

For an integer $N\geq1$, we write $[N]:=\{1,\ldots,N\}$. Unless stated otherwise, all matrices have dimension $2^n\times2^n$. For a matrix $A$, we denote its $(i,j)$-th entry by~$A(i,j)$, its $i$-th row by~$A(i,\cdot)$, and its spectrum by $\sigma(A)$. 
We write $\|A\|_2$ for the operator norm and use the induced matrix norms
\[
    \|A\|_1:=\max_{j}\sum_i |A(i,j)|,
        \qquad \|A\|_\infty:=\max_i\sum_j |A(i,j)|.
\]
Thus, $\|A\|_1$ denotes the maximum absolute column sum, not the trace norm.
If $A$ is Hermitian, then $\|A\|_1=\|A\|_\infty$. Further,
\[
    \|A(i,\cdot)\|_1\leq \|A\|_1
        \qquad\text{for every }i\in[2^n].
\]
Moreover, if every column of $A$ contains at most $s$ nonzero entries, then Cauchy-Schwarz gives
\[
    \|A\|_1\leq\sqrt{s}\,\|A\|_2.
\]

For a polynomial $p(x)=\sum_{k=0}^d c_kx^k$, we write
\[
    \|p\|_1 := \sum_{k=0}^d |c_k|
\]
for its monomial-basis coefficient $\ell_1$-norm.

For a Hermitian matrix with spectral decomposition $A=\sum_{k=1}^{2^n}\lambda_k v_k v_k^\dagger$ and a scalar function~$f$ defined on $\sigma(A)$, we use the standard functional calculus
\[
    f(A):=\sum_{k=1}^{2^n} f(\lambda_k) v_kv_k^\dagger.
\]\vspace{-.2cm}
In particular,\vspace{-.1cm}
\[
    \tr[f(A)] = \sum_{k=1}^{2^n} f(\lambda_k)
        \qquad \text{and} \qquad
            \frac{1}{2^n}\tr[f(A)] = \frac{1}{2^n}\sum_{k=1}^{2^n} f(\lambda_k).
\]
We refer to these quantities as the unnormalized and normalized spectral sum, respectively.

\subsection{Probabilistic and analytic tools}

We begin with the standard concentration inequality due to Hoeffding.
\begin{fact}[Hoeffding's inequality \cite{Hoe63}]
    Let $X_1,...,X_T$ be independent random variables with expectation $\mathbb{E}[X_i]=\mu$ such that $a\leq X_i \leq b$ and let $\overline{X}=\frac{1}{T}(X_1+...+X_T)$.
    Then, for all $\varepsilon>0$,
    \[
        \mathrm{Pr}\!\left[|\overline{X}-\mu|\geq \varepsilon\right] 
            \leq 2\exp\!\left(\frac{-2T\varepsilon^2}{(b-a)^2}\right).
    \]
\end{fact}

We will also use Taylor's theorem. Let us denote by $f^{(k)}$ the $k$-th derivative of $f$.
\begin{fact}[Taylor's theorem]
\label{fact:taylor}
    Let $I\subseteq\mathbb{R}$ be an interval, $x,x_0\in I$ and $f : I \rightarrow \mathbb{R}$ $n$ times differentiable.
    Then there exists a $\xi$ between $x$ and $x_0$ such that
    \[
        f(x) = \sum_{k=0}^{n-1} \frac{f^{(k)}(x_0)}{k!} (x-x_0)^k + \frac{f^{(n)}(\xi)}{n!} (x-x_0)^n.
    \]
\end{fact}

\subsection{Complexity classes}

We next recall the complexity classes used in our hardness and containment results.

\begin{definition}($\DQC 1$ and $\DQC k$)
\label{def:DQC1}
    Let $L=(L_{\mathrm{yes}},L_{\mathrm{no}})$ be a promise problem.
    Then $L\in \DQC k$ if and only if there exist functions $a,b:\N\rightarrow[0,1]$ such that $b-a \geq 1/\poly(n)$ and a polynomial-time generated family of quantum circuits $Q = \{Q_x ~|~ x\in L_{\mathrm{yes}}\cup L_{\mathrm{no}}\}$, where each circuit $Q_x$ acts on $n=\poly(|x|)$ input qubits with initial state $\rho:=\ket{0}\bra{0}^{\otimes k}\otimes \frac{1}{2^{n-k}}I_{n-k}$ and consists of~$T=\poly(|x|)$ gates such that
    \begin{itemize}
        \item if $x\in L_{\mathrm{yes}}$,
        \[
            \mu_{\mathrm{accept}}=\tr \left[(\ket{1}\bra{1}^{\otimes 1}\otimes I_{n-1})Q_x\rho Q_x^\dagger \right] \geq b,
        \]
        \item if $x\in L_{\mathrm{no}}$,
        \[
            \mu_{\mathrm{accept}}=\tr \left[(\ket{1}\bra{1}^{\otimes 1}\otimes I_{n-1})Q_x\rho Q_x^\dagger \right] \leq a.
        \]
    \end{itemize}
\end{definition}

We use the standard sampling-based convention that the circuit may be executed polynomially many times on fresh copies of its initial state. Thus an inverse-polynomial completeness-soundness gap can be resolved with bounded error.
In contrast, \cite{FKM+16} introduced a variant called~$\mathsf{BQ_{[\mathit{k}]}P}$, where the completeness-soundness gap is constant ($b-a=\Omega(1)$), so that a single run suffices, and the number of \emph{total} clean qubits needed is indeed $k$.
The following robustness property is well known, see e.g. \cite{SJ08} for a proof.

\begin{fact}
    $\DQC1 = \DQC k$ for $k=O(\log n)$.
\end{fact}

\begin{definition}[$\PP$]
    Let $L=(L_{\mathrm{yes}},L_{\mathrm{no}})$ be a promise problem. Then $L \in \PP$ if and only if there exists a polynomial-time probabilistic Turing machine such that
    \begin{itemize}
        \item if $x\in L_{\mathrm{yes}}$, the Turing machine accepts with probability strictly larger than $\frac{1}{2}$, and
        \item if $x\in L_{\mathrm{no}}$, the Turing machine accepts with probability strictly smaller than $\frac{1}{2}$.
    \end{itemize}
\end{definition}

\subsection{Matrix-access models}

We now describe the input models for matrices used throughout the paper. All matrix entries are assumed to have bit complexity polynomial in $n$.

\begin{definition}[Query access to matrices]
    We say that we have query access to a matrix~$A~\in~\C^{2^n \times 2^n}$ in time $Q_A$ if we have access to a classical algorithm $\mathcal{Q}_A$ which on input~$(i,j)~\in~[2^n]~\times~[2^n]$, returns entry $A(i,j)$ in time $Q_A$.
\end{definition}
\begin{definition}[Sparse access to matrices]
    We say that we have sparse access to an $s$ sparse matrix $A\in\C^{2^n\times2^n}$ in time $Q_A$ if we have query-access to $A$ and there exist two more classical algorithms $\mathcal{Q}_A^\mathrm{row}$ and $\mathcal{Q}_A^\mathrm{col}$ running in time $Q_A$ such that
    \begin{itemize}
        \item on inputs $(i,l) \in [2^n] \times [s]$, the algorithm $\mathcal{Q}_A^\mathrm{row}$ outputs the $l$-th non-zero entry of the $i$-th row of $A$ and its corresponding column index if this row has at least $l$ non-zero entries, and outputs an error message otherwise,
        \item on inputs $(j,l) \in [2^n] \times [s]$, the algorithm $\mathcal{Q}_A^\mathrm{col}$ outputs the $l$-th non-zero entry of the $j$-th column of $A$ and its corresponding row index if this column has at least $l$ non-zero entries, and outputs an error message otherwise.
    \end{itemize}
    Unless stated otherwise, we always assume that the query-access time $Q_A$ is polynomial in $n$.
\end{definition}
\noindent 

For Hermitian matrices, row and column access are equivalent.
\begin{definition}[$k$-local Hamiltonian]\label{def:local-Hamiltonian}
    A $k$-local Hamiltonian acting on $n$ qubits is a Hermitian matrix
    \[
        H=\sum_{i=1}^m H_i\in\mathbb{C}^{2^n\times2^n},
    \]
    where $m=\poly(n)$ and every term $H_i$ acts nontrivially on at most $k$ qubits. The input specifies, for each term, the qubits on which it acts and provides query access to its local matrix.
    We call~$H$~\emph{log-local} if~$k=O(\log n)$.
\end{definition}

Each $k$-local term is $2^k$-sparse, and hence a sum of $m$ such terms is at most $m2^k$-sparse. In particular, every log-local Hamiltonian with polynomially many terms is polynomially sparse. For a fixed local decomposition, we call
\[
    \Gamma(H):=\sum_{i=1}^m\|H_i\|_2
\]
the \emph{total interaction strength}. In our classical algorithms for local Hamiltonians, we normalize the decomposition so that $\Gamma(H)\leq1$.

We generally write $A$ for a matrix given by sparse access and $H$ for a local Hamiltonian.


\section{Two simple classical algorithms for spectral sum estimation}\label{sec:classical_algorithm}

In this section, we develop two classical algorithms for estimating normalized spectral sums. The first evaluates diagonal entries of matrix polynomials by deterministically enumerating sparse walks. The second replaces this enumeration by a random walk and is particularly useful for local Hamiltonians with bounded total interaction strength.

After proving general guarantees for the two estimators, we combine them with polynomial approximations of the logarithm, exponential, monomial, and reciprocal functions. This yields classical algorithms for the log-determinant, partition function, and traces of powers and inverses.

Readers interested only in the algorithm for sparse matrices may skip \cref{alg:stochastic-sparse-walk} and the discussion following it.

\subsection{An algorithm based on deterministic sparse matrix powering}
\label{subsec: simple classical algorithm}

Let $p(x)=\sum_{k=0}^d c_kx^k$ approximate the target function $f$. Since 
\[ 
    \frac{1}{2^n}\tr[p(A)] 
        = \mathbb{E}_{i\in[2^n]}[p(A)(i,i)] 
        \qquad\text{and}\qquad 
            p(A)(i,i)=\sum_{k=0}^d c_kA^k(i,i), 
\] 
it suffices to sample a uniformly random index $i$ and compute the diagonal entries $A^k(i,i)$, as explained in the introduction.
For an $s$-sparse matrix, these are weighted sums over closed walks starting and ending at $i$, which can be evaluated by exhaustive enumeration.

\begin{lemma}[Deterministic sparse powering]
\label{lem: deterministic sparse powering}
    Given sparse access to an $s$-sparse matrix \mbox{$A\in\mathbb{C}^{2^n\times 2^n}$}, an index $i\in[2^n]$, and an integer $d\geq0$, there is a deterministic classical algorithm running in time $O^*(s^d)$ which outputs all values
    \[
        A^0(i,i),A^1(i,i),\dots,A^d(i,i).
    \]
\end{lemma}

\begin{proof}
    The algorithm enumerates all nonzero walks
    \[
        i=i_0\to i_1\to\cdots\to i_k
    \]
    of length $k\leq d$ starting at $i$. For each such walk, it maintains its weight
    \[
        A(i,i_1)A(i_1,i_2)\cdots A(i_{k-1},i_k).
    \]
    Whenever the walk returns to $i$ at time $k$, i.e. $i_k=i$, this weight is added to the accumulator for $A^k(i,i)$. Therefore, the accumulator for length $k$ equals
    \[
        \sum_{i_1,\dots,i_{k-1}} A(i,i_1)A(i_1,i_2)\cdots A(i_{k-1},i)
            = A^k(i,i).
    \]
    Since every vertex has at most $s$ nonzero neighbors, the number of enumerated walks of length at most $d$ is bounded by
    \[
        1+s+s^2+\cdots+s^d=O^*(s^d).\qedhere
    \]
\end{proof}

Combining \cref{lem: deterministic sparse powering} with uniform sampling of the diagonal gives \cref{alg:deterministic-sparse-powering}. Each sample is exactly one diagonal entry of $p(A)$, and the final estimate is their empirical average.

\begin{algorithm}[H]
\caption{Deterministic sparse-powering based estimator}
\label{alg:deterministic-sparse-powering}
\begin{algorithmic}[1]
\Input sparse access to a Hermitian matrix $A\in\mathbb{C}^{2^n\times 2^n}$, a real polynomial $p(x)=\sum_{k=0}^d c_kx^k$, accuracy $\varepsilon>0$, failure probability $\delta>0$.
\Output An $\varepsilon$-additive approximation of $2^{-n}\tr[p(A)]$ with probability at least $1-\delta$.

\State Set $T\gets \left\lceil 2\|p(A)\|_2^2\log(2/\delta)/\varepsilon^2\right\rceil$

\For{$t=1,\dots,T$}
    \State Sample $i\in[2^n]$ uniformly at random
    \State Compute $(A^0(i,i),A^1(i,i),\dots,A^d(i,i))$ deterministically using
    \cref{lem: deterministic sparse powering}
    \State Set $Y_t\gets \sum_{k=0}^d c_k A^k(i,i)$
\EndFor

\State \Return $\widehat{\mu}:=\frac{1}{T}(\Re[Y_1] + \dots +\Re[Y_T])$
\end{algorithmic}
\end{algorithm}

The next lemma records the resulting accuracy and running time.

\begin{lemma}
\label{lem: deterministic powering trace of polynomial}
    Let $A\in\mathbb{C}^{2^n\times 2^n}$ be an $s$-sparse Hermitian matrix given by sparse access, and let~$p(x)=\sum_{k=0}^d c_kx^k$ be a degree~$d$ real polynomial. Then,~\cref{alg:deterministic-sparse-powering} outputs with probability at least $1-2^{-\poly(n)}$ an $\varepsilon$-additive estimate
    \[
        \frac{1}{2^n}\tr[p(A)] \pm \varepsilon
            \qquad\text{in time}\qquad
                O^*\!\left(s^d\|p(A)\|_2^2/\varepsilon^2\right).
    \]
\end{lemma}
\begin{proof}
    We prove that one iteration of~\cref{alg:deterministic-sparse-powering} gives an unbiased estimator for $2^{-n}\tr[p(A)]$. Let $i\in[2^n]$ be sampled uniformly at random. By \cref{lem: deterministic sparse powering}, the algorithm computes
    \[
        Y_t = \sum_{k=0}^d c_kA^k(i,i)
            = p(A)(i,i).
    \]
    Therefore,
    \[
        \mathbb{E}[Y_t]
        =
        \frac{1}{2^n}\sum_{i=1}^{2^n}p(A)(i,i)
        =
        \frac{1}{2^n}\tr[p(A)].
    \]
    Since $A$ is Hermitian and $p$ is a real polynomial, $p(A)$ is Hermitian,
    so $\tr[p(A)]$ is real. Thus~$\Re[Y_t]$ is also an unbiased estimator of
    $2^{-n}\tr[p(A)]$.

    We next bound the range of the estimator. For every $i$,
    \[
        |Y_t|
        =
        |p(A)(i,i)|
        \leq
        \|p(A)\|_2
    \]
    Hoeffding's inequality applied to the $T$ independent samples
    $\Re[Y_1],\dots,\Re[Y_T]$ gives
    \[
        \mathrm{Pr}\!\left[
            \left|
                \frac1T\sum_{t=1}^T\Re[Y_t]
                -
                \frac{1}{2^n}\tr[p(A)]
            \right|
            \geq \varepsilon
        \right]
        \leq
        2\exp\!\left(
            -\frac{2T\varepsilon^2}{4\|p(A)\|_2^2}
        \right).
    \]
    With
    \[
        T=
        \left\lceil
            2\|p(A)\|_2^2\log(2/\delta)/\varepsilon^2
        \right\rceil,
    \]
    this failure probability is at most $\delta$.

    Finally, by \cref{lem: deterministic sparse powering}, computing one
    sample takes time $O^*(s^d)$. Hence, with $\delta \geq 2^{-\poly(n)}$, the total runtime is
    \[
        O^*(s^dT)
            = O^*\!\left( s^d\|p(A)\|_2^2/\varepsilon^2 \right).\qedhere
    \]
\end{proof}

We now pass from matrix polynomials to general spectral functions. If $p$ uniformly approximates $f$ on the spectrum of $A$, then their normalized spectral sums differ by at most the same uniform error. Combining this observation with \cref{lem: deterministic powering trace of polynomial} gives our first general spectral-sum theorem.

\begin{theorem}
\label{thm: deterministic powering trace of general function}
    Let $I\subseteq \mathbb{R}$ and let $f: I \rightarrow \mathbb{R}$ be $\varepsilon$-approximated by a polynomial-time computable real polynomial~$p(x)$ of degree $d$, i.e.
    \[
        \sup_{x\in I} \left| f(x) - p(x) \right|
            \leq \varepsilon.
    \]
    Let $A\in\mathbb{C}^{2^n \times 2^n}$ be an $s$-sparse Hermitian matrix with spectrum in $I$ given by sparse access.
    Then, there is a classical algorithm that outputs with probability at least $1-2^{-\poly(n)}$ a $(2\varepsilon)$-additive approximation of the normalized spectral sum
    \[
        \frac{1}{2^n}\tr[f(A)]\pm 2\varepsilon
            \qquad \text{in time} \qquad
                O^*\left( s^d f_{\max}^2/\varepsilon^2 \right),
    \]
    where $f_{\max}$ is the supremum of $|f|$ on $I$.
\end{theorem}
\begin{proof} 
    Observe that, 
    \[ 
        \left| \frac{1}{2^n}\tr[f(A)] - \frac{1}{2^n}\tr[p(A)] \right| 
            \leq \frac{1}{2^n}\sum_{i=1}^{2^n} |f(\lambda_i) - p(\lambda_i)|
            \leq \varepsilon.
    \]
    Moreover, we can assume that $\varepsilon\leq f_{\max}$ as the statement becomes trivial otherwise.
    Then,
    \[ 
        \|p(A)\|_2^2 
            \leq \sup_{x\in I}|p(x)|^2
            \leq \left(f_{\max}+\varepsilon\right)^2
            =O\!\left(f_{\max}^2\right).
    \] 
    The result now follows by applying \cref{lem: deterministic powering trace of polynomial} to estimate $2^{-n}\tr[p(A)]$ to additive accuracy~$\varepsilon$.
\end{proof}

\subsection{An algorithm based on random walks}

The deterministic estimator pays the full cost of enumerating all walks of length at most $d$. We next replace this exhaustive search by a random walk. From a current index $i$, the algorithm chooses a nonzero entry $A(i,j)$ with probability proportional to $|A(i,j)|$. It compensates for this choice by multiplying the running weight by the row $1$-norm and the complex sign of the sampled entry. Consequently, the expected contribution of a sampled closed walk is exactly its matrix-product weight.

A single trajectory of length $d$ simultaneously provides unbiased estimators for $A^k(i,i)$ for every $k\leq d$: whenever the walk returns to its starting point at time $k$, the current weight contributes to the coefficient $c_k$. The resulting estimator is described below.

\begin{algorithm}[H]
\caption{Random walk based estimator}
\label{alg:stochastic-sparse-walk}
\begin{algorithmic}[1]
\Input sparse access to a Hermitian matrix $A\in\mathbb{C}^{2^n\times 2^n}$,
a real polynomial $p(x)=\sum_{k=0}^d c_kx^k$, accuracy $\varepsilon>0$, failure probability $\delta>0$.
\Output An $\varepsilon$-additive approximation of $2^{-n}\tr[p(A)]$ with probability at least $1-\delta$.

\State Set $T\gets \left\lceil 2\max\{1,\|A\|_1^{2d}\}\|p\|_1^2\log(2/\delta)/\varepsilon^2\right\rceil$

\For{$t=1,\dots,T$}
    \State Set $Y_t\gets c_0$ and $W\gets 1$
    \State Sample $X_0=i\in[2^n]$ uniformly at random.
    \For{$k=1,\dots,d$}
        \If{$\|A(X_{k-1},\cdot)\|_1=0$}
            \State \textbf{break}
        \EndIf
        \State Sample $X_k$ from the nonzero entries of $A(X_{k-1},\cdot)$ with probability
        \vspace{-.2cm}
        \[
            \mathrm{Pr}[X_k=j\mid X_{k-1}] 
                = \frac{|A(X_{k-1},j)|}{\|A(X_{k-1},\cdot )\|_1}
        \]
        \State Set $W\gets W\cdot\operatorname{sgn}(A(X_{k-1},X_k)) \cdot \|A(X_{k-1},\cdot)\|_1$
        \If{$X_k=X_0$}
            \State Set $Y_t\gets Y_t+c_kW$
        \EndIf
    \EndFor
\EndFor

\State \Return $\widehat{\mu}:=\frac{1}{T}(\Re[Y_1] + \dots + \Re[Y_T])$
\end{algorithmic}
\end{algorithm}

The analysis has two parts. First, the importance weight exactly cancels the transition probabilities, making the estimator unbiased. Second, its magnitude is bounded in terms of~$\|A\|_1^d\|p\|_1$, which allows us to apply Hoeffding's inequality.

\begin{lemma}
\label{lem: random-walk trace of polynomial}
    Let $A\in\mathbb{C}^{2^n\times 2^n}$ be an $s$-sparse Hermitian matrix given via sparse access, and let~$p(x)=\sum_{k=0}^d c_kx^k$ be a degree~$d$ real polynomial. Then,~\cref{alg:stochastic-sparse-walk} outputs, with probability at least $1-\delta$, an additive estimate
    \[
        \frac{1}{2^n}\tr[p(A)]\pm\varepsilon
            \qquad\text{in time}\qquad
                O^*\!\left(sd\max\{1,\|A\|_1^{2d}\}\|p\|_1^2\log(1/\delta)/\varepsilon^2\right).
    \]
    In particular, when $s,d\leq \poly(n)$ and $\delta\geq2^{-\poly(n)}$, the running time is
    \[
        O^*\!\left(\max\{1,\|A\|_1^{2d}\}\|p\|_1^2/\varepsilon^2\right).
    \]
\end{lemma}
\begin{proof}
    We prove that each $Y_t$ in~\cref{alg:stochastic-sparse-walk} gives an unbiased estimator for $\frac{1}{2^n}\tr[p(A)]$.
    Fix the initial index $X_0=i_0$.
    Set $W_0 = 1$ and for each $k\in\{1,\dots,d\}$, let $W_k$ denote the value of the running weight $W$ after $k$ random walk steps, if no zero row is encountered during the walk.
    Thus, if no zero row is encountered, we find that
    \[
        W_k = \prod_{\ell=0}^{k-1} \operatorname{sgn}(A(X_\ell,X_{\ell+1})) \cdot \|A(X_\ell,\cdot)\|_1,
    \]
    where $\operatorname{sgn}(A(i,j)):=\frac{A(i,j)}{|A(i,j)|}$ denotes the complex sign of an entry.
    If the walk reaches a zero row before time $k$, then the algorithm aborts the walk.
    This agrees with the fact that no zero row can ever contribute to $A^k(i_0,i_0)$.

    For $k=0$, we have $W_0=1=A^0(i_0,i_0)$, and for $k\geq1$, the transition probability of a path~$i_0,i_1,\dots,i_k$ is
    \[
        \prod_{\ell=0}^{k-1}
        \frac{|A(i_\ell,i_{\ell+1})|}{\|A(i_\ell,\cdot)\|_1}.
    \]
    Therefore,
    \begin{align*}
        \mathbb{E}\!\left[
            \mathbbm{1}_{[X_k=X_0]}W_k
            \,\middle|\,
            X_0=i_0
        \right]
        &=
        \sum_{i_1,\dots,i_k}
        \mathbbm{1}_{[i_k=i_0]}
        \prod_{\ell=0}^{k-1}
        \left(
            \operatorname{sgn}(A(i_\ell,i_{\ell+1}))
            \|A(i_\ell,\cdot)\|_1
            \frac{|A(i_\ell,i_{\ell+1})|}
                 {\|A(i_\ell,\cdot)\|_1}
        \right) \\
        &=
        \sum_{i_1,\dots,i_k}
        \mathbbm{1}_{[i_k=i_0]}
        \prod_{\ell=0}^{k-1}
        A(i_\ell,i_{\ell+1}) \\
        &=
        A^k(i_0,i_0).
    \end{align*}

    The value of $Y_t$ produced in each iteration of the algorithm is
    \[
        Y_t = c_0 + \sum_{k=1}^d c_k\, \mathbbm{1}_{[X_k=X_0]}W_k.
    \]
    Hence, conditioning on $X_0=i_0$,
    \[
        \mathbb{E}[Y_t\mid X_0=i_0]
            = \sum_{k=0}^d c_k A^k(i_0,i_0)
            = p(A)(i_0,i_0).
    \]
    Averaging over the uniformly random choice of $X_0$ gives
    \[
        \mathbb{E}[Y_t]
            = \frac{1}{2^n} \sum_{i=1}^{2^n}p(A)(i,i)
            = \frac{1}{2^n}\tr[p(A)].
    \]
    Since $A$ is Hermitian and $p$ has real coefficients, $p(A)$ is also Hermitian, so $\tr[p(A)]$ is real. 
    Therefore~$\Re[Y_t]$ is an unbiased estimator of $\frac{1}{2^n}\tr[p(A)]$.

    We next bound the range of the estimator. For every row index $i$, $\|A(i,\cdot)\|_1\leq \|A\|_1$.
    Therefore, for every $k$, $|W_k|\leq \|A\|_1^k$, and consequently,
    \[
        |Y_t|
        \leq
        \sum_{k=0}^d |c_k|\,\|A\|_1^k
        \leq
        \|p\|_1\max\{1,\|A\|_1^d\},
    \]
    where the maximum handles the case where $\|A\|_1 < 1$.
    Hoeffding's inequality applied to the $T$ independent samples
    $\Re[Y_1],\dots,\Re[Y_T]$ now gives
    \[
        \mathrm{Pr}\!\left[
            \left|
                \frac1T\sum_{t=1}^T \Re[Y_t]
                -
                \frac{1}{2^n}\tr[p(A)]
            \right|
            \geq \varepsilon
        \right]
        \leq
        2\exp\!\left(
            -\frac{2T\varepsilon^2}
                  {4\max\{1,\|A\|_1^{2d}\}\|p\|_1^2}
        \right).
    \]
    With
    \[
        T = \left\lceil 2\max\{1,\|A\|_1^{2d}\}\|p\|_1^2\log(2/\delta)/\varepsilon^2\right\rceil,
    \]
    this failure probability is at most $\delta$.

    Finally, each random-walk step can be implemented in time $O^*(s)$ using sparse access to the rows of $A$, and each sample uses at most $d$ steps. Thus the total runtime is
    \[
        O^*(s d T) = O^*\!\left(\max\{1,\|A\|_1^{2d}\}\|p\|_1^2\log(1/\delta)/\varepsilon^2\right).
    \]
    This proves the theorem.
\end{proof}

A priori, the runtime bounds of the algorithms, being
\[
    O^*\left(s^d\|p(A)\|_2^2/\varepsilon^2\right)
        \qquad\text{and}\qquad
            O^*\left(\max\{1,\|A\|_1^{2d}\}\|p\|_1^2/\varepsilon^2\right),
\]
are incomparable.
However, recall that for an $s$-sparse Hermitian matrix, $\|A\|_1\leq\sqrt{s}\,\|A\|_2$, so when $\|A\|_2\leq1$ the random-walk estimator has at most the same $s^d$ dependence, but additionally depends on $\|p\|_1$. It can however significantly outperform the sparse powering based algorithm when $\|A\|_1$ is much smaller than the generic bound~$\sqrt{s}\|A\|_2$.

This occurs for local Hamiltonians with bounded total interaction strength. If
\[
    H=\sum_{i=1}^m H_i,
        \qquad
    \sum_{i=1}^m\|H_i\|_2\leq1,
\]
and every $H_i$ acts nontrivially on at most $k$ qubits, then each $H_i$ is $2^k$-sparse and
\[
    \|H\|_1
        \leq\sum_{i=1}^m\|H_i\|_1
        \leq2^{k/2}\sum_{i=1}^m\|H_i\|_2
        \leq2^{k/2}.
\]
Consequently, $\|H\|_1^{2d}\leq2^{kd}$, independently of the number of local terms $m$. Treating $H$ only as a sparse matrix would instead give sparsity at most $m2^k$ and a factor $(m2^k)^d$. This observation yields the following theorem.

\begin{theorem}
\label{thm: random walk trace of function for Hamiltonians}
    Let $I\subseteq \mathbb{R}$ and let $f: I \rightarrow \mathbb{R}$ be $\varepsilon$-approximated by a polynomial-time computable real polynomial~$p(x)=\sum_{k=0}^d c_k x^k$ of degree $d$, i.e.
        \[
            \sup_{x\in I} \left| f(x) - p(x) \right|
                \leq \varepsilon.
        \]
    Let $H = \sum_{i=1}^m H_i$ be a $k$-local Hamiltonian with spectrum in $I$, acting on $n$ qubits, and with bounded total interaction strength 
    \[
        \sum_{i=1}^m \|H_i\|_2 \leq 1.
    \]
    Then, \cref{alg:stochastic-sparse-walk} run on $p$, outputs with probability at least~\mbox{$1-2^{-\poly(n)}$}, an additive estimate
    \[
        \frac{1}{2^n}\tr[f(H)]\pm2\varepsilon
            \qquad\text{in time}\qquad
                O^*\!\left(2^{kd}\|p\|_1^2/\varepsilon^2\right).
    \]
\end{theorem}
\begin{proof}
    As before, the uniform approximation guarantee implies
    \[
        \left|\frac{1}{2^n}\tr[f(H)]-\frac{1}{2^n}\tr[p(H)]\right|
            \leq\varepsilon.
    \]
    Apply \cref{lem: random-walk trace of polynomial} to estimate $2^{-n}\tr[p(H)]$ to additive error $\varepsilon$. Since $\|H\|_1\leq2^{k/2}$, its running time indeed is
    \[
        O^*\!\left(2^{kd}\|p\|_1^2/\varepsilon^2\right).\qedhere
    \]
\end{proof}

\subsection{Applications}

We now combine the two general algorithms with polynomial approximations of the logarithm, exponential, monomial, and reciprocal functions. The approximation results are proved in the appendix: the exponential and monomial approximations are taken from \cite{SV14}, with their coefficient-norm bounds made explicit, while the logarithm and reciprocal are obtained by Chebyshev truncation.
The resulting polynomials have the degrees shown in the second column of \cref{tab: runtimes} and, in every case, their monomial-basis coefficient norms are bounded exponentially in their degree~\mbox{$\|p\|_1=\exp(O(d))$}, see \cref{lem: log-approx,lem: exp-approx,lem: monomial-approx,lem: inv-approx}.

\begin{corollary}[Classical algorithms for spectral sums]
\label{cor: classical applications}
    Let $\varepsilon\in(0,1/2)$. There is a universal constant $c>1$ such that every quantity in the first column of \cref{tab: runtimes} can be estimated to additive error $\varepsilon$ with probability at least $1-2^{-\poly(n)}$ as follows.
    \begin{enumerate}
        \item If $M=A$ is an $s$-sparse Hermitian matrix, where $s\geq2$, given by sparse access, then the running time is given in the third column.
        \item If $M=H=\sum_{i=1}^m H_i$ is a $k$-local Hamiltonian satisfying
        \[
            \sum_{i=1}^m\|H_i\|_2\leq1,
        \]
        then the running time is given in the fourth column.
    \end{enumerate}
\end{corollary}

\begin{table}[H]
\centering
\small
\renewcommand{\arraystretch}{1.4}
\setlength{\tabcolsep}{3.5pt}
\resizebox{\textwidth}{!}{
\begin{tabular}{@{}c|c|c|c@{}}
\hline
Spectral sum and promise
& Approximation degree $d$
& Runtime for $s$-sparse matrices
& Runtime for $k$-local Hamiltonians \\
\hline
    \begin{tabular}{@{}c@{}}
        $\frac{1}{2^n}\log\det(M)\pm\varepsilon$\\
        $\sigma(M)\subseteq[1/\kappa,1]$
    \end{tabular}
    & $O\!\left(\sqrt{\kappa}\log(1/\varepsilon)\right)$
    & $O^*\!\left(s^{c\sqrt{\kappa}\log(1/\varepsilon)}\right)$
    & $O^*\!\left(2^{ck\sqrt{\kappa}\log(1/\varepsilon)}\right)$ \\
    \hline
    \begin{tabular}{@{}c@{}}
        $\frac{1}{2^n}\tr[e^{-\beta M}]\pm\varepsilon$\\
        $\sigma(M)\subseteq[0,1]$
    \end{tabular}
    & $O\!\left(\sqrt{\beta\log(1/\varepsilon)}+\log(1/\varepsilon)\right)$
    & $O^*\!\left(s^{c(\sqrt{\beta\log(1/\varepsilon)}+\log(1/\varepsilon))}\right)$
    & $O^*\!\left(2^{ck(\sqrt{\beta\log(1/\varepsilon)}+\log(1/\varepsilon))}\right)$ \\
    \hline
    \begin{tabular}{@{}c@{}}
        $\frac{1}{2^n}\tr[M^p]\pm\varepsilon$\\
        $\sigma(M)\subseteq[-1,1]$
    \end{tabular}
    & $O\!\left(\sqrt{p\log(1/\varepsilon)}\right)$
    & $O^*\!\left(s^{c(\sqrt{p\log(1/\varepsilon)}+\log(1/\varepsilon))}\right)$
    & $O^*\!\left(2^{ck(\sqrt{p\log(1/\varepsilon)}+\log(1/\varepsilon))}\right)$ \\
    \hline
    \begin{tabular}{@{}c@{}}
        $\frac{1}{2^n}\tr[M^{-1}]\pm\varepsilon$\\
        $\sigma(M)\subseteq[1/\kappa,1]$
    \end{tabular}
    & $O\!\left(\sqrt{\kappa}\log(\kappa/\varepsilon)\right)$
    & $O^*\!\left(s^{c\sqrt{\kappa}\log(\kappa/\varepsilon)}\right)$
    & $O^*\!\left(2^{ck\sqrt{\kappa}\log(\kappa/\varepsilon)}\right)$ \\
    \hline
    \end{tabular}
}
\caption{Polynomial-approximation degrees and the resulting runtimes for the two classical algorithms.}
\label{tab: runtimes}
\end{table}

\begin{proof}
    Use polynomial approximation error $\varepsilon/2$ and estimation error $\varepsilon/2$. 
    For sparse matrices, apply \cref{thm: deterministic powering trace of general function}. The values of $f_{\max}$ for the four rows are $\log\kappa$, $1$, $1$, and $\kappa$.
    Since $s\geq2$ and~$\varepsilon<1/2$, the resulting $f_{\max}^2/\varepsilon^2$ factors can be absorbed into the displayed exponents. 
    For local Hamiltonians, apply \cref{thm: random walk trace of function for Hamiltonians}. In each case $\|p\|_1=\exp(O(d))$, so the coefficient-norm and~$\varepsilon^{-2}$ factors can likewise be absorbed by increasing the universal constant $c$.
\end{proof}

In our setting, we always have $\|M\|_2\leq1$. Thus, the random-walk estimator actually gives the same asymptotic runtime bounds also for sparse matrices.
We used the deterministic sparse powering based estimator above only for its simplicity.

For the log-determinant, an additional spectral upper bound converts additive accuracy for the normalized quantity into relative accuracy for the unnormalized quantity.

\begin{corollary}
\label{cor: relative error classical logdet approx}
    Let $s,\kappa\geq2$ and $\varepsilon\in(0,1/2)$, and suppose that we are
    given sparse access to an $s$-sparse Hermitian matrix
    $A\in\mathbb{C}^{2^n\times2^n}$ with
    \[
        \sigma(A)\subseteq[1/\kappa,1-1/\kappa].
    \]
    Then there is a classical algorithm which, with probability at least
    $1-2^{-\poly(n)}$, outputs
    \[
        \log\det(A)\pm\varepsilon|\log\det(A)|
        \qquad\text{in time}\qquad
        O^*\!\left(s^{c\sqrt{\kappa}\log(\kappa/\varepsilon)}\right),
    \]
    where $c>1$ is a universal constant.
\end{corollary}

\begin{proof}
    Set $\mu:=2^{-n}\log\det(A)$. Since
    $\log(1-1/\kappa)\leq-1/\kappa$, the spectral promise gives
    $\mu\leq-1/\kappa$ and hence $|\mu|\geq1/\kappa$. Apply
    \cref{cor: classical applications} with additive accuracy
    $\eta:=\varepsilon/\kappa$. If $z$ is the resulting estimate of $\mu$,
    then
    \[
        |2^n z-\log\det(A)|
            \leq2^n\eta
            \leq\varepsilon|\log\det(A)|.
    \]
    Substituting $\eta=\varepsilon/\kappa$ into the first sparse-matrix
    runtime in \cref{tab: runtimes} proves the claim.
\end{proof}


\section{DQC1-completeness of estimating normalized spectral sums}\label{sec:DQC1}

\subsection{DQC1-containment}

We begin with the following uniform formulation of a result of Cade and Montanaro \cite{CM17}.

\begin{lemma}[cf. Corollary~6 in \cite{CM17}]
\label{lem: CM17 Lipschitz}
    Let $I_n\subseteq\mathbb{R}$ and let $f_n:I_n\rightarrow\mathbb{R}$ be a uniformly logspace-computable family of functions.
    Suppose that each $f_n$ is Lipschitz continuous on $I_n$ with Lipschitz constant $K_n$, and write
    \[
        M_n:=\sup_{x\in I_n}|f_n(x)|.
    \]
    Then, for every log-local Hamiltonian $H\in\mathbb{C}^{2^n\times2^n}$ with $\sigma(H)\subseteq I_n$, there is a $\DQC1$ algorithm which estimates
    \[
        \frac{1}{2^n}\tr[f_n(H)] \pm \varepsilon(K_n+M_n),
    \]
    for every $\varepsilon=\Omega(1/\poly(n))$.
\end{lemma}

In particular, whenever $K_n+M_n=\poly(n)$, the normalized spectral sum can be estimated to inverse-polynomial additive accuracy in $\DQC1$. Applying this observation to the four functions considered in this paper gives the following corollary.

\begin{corollary}
\label{cor: spectral sums in DQC1}
    Let $H\in\mathbb{C}^{2^n\times2^n}$ be a log-local Hamiltonian.
    The following quantities can be estimated to inverse-polynomial additive accuracy in $\DQC1$:
    \begin{enumerate}
        \item the normalized log-determinant $\frac{1}{2^n}\log\det(H)$,
            provided $\sigma(H)\subseteq[1/\kappa,1]$ and $\kappa=\poly(n)$.
        \item the normalized partition function $\frac{1}{2^n}\tr[e^{-\beta H}]$, provided $\sigma(H)\subseteq[0,1]$ and $\beta=\poly(n)$.
        \item the normalized trace of the $p$-th power $\frac{1}{2^n}\tr[H^p]$, provided $\sigma(H)\subseteq[-1,1]$ and $p=\poly(n)$.
        \item the normalized trace of the inverse $\frac{1}{2^n}\tr[H^{-1}]$, provided $\sigma(H)\subseteq[1/\kappa,1]$ and $\kappa=\poly(n)$.
    \end{enumerate}
\end{corollary}

While the application of \cref{lem: CM17 Lipschitz} to obtain the corollary is rather direct, and already explicitly stated in case of the trace of a polynomial power in \cite{CM17}, we give the short argument here.

\begin{proof}
    We apply \cref{lem: CM17 Lipschitz}. The required bounds are summarized below:
    \[
    \begin{array}{c|c|c|c}
        f(x) & I & K & \sup_{x\in I}|f(x)| \\ \hline
        \log x
            & [1/\kappa,1]
            & \kappa
            & \log\kappa \\[2mm]
        e^{-\beta x}
            & [0,1]
            & \beta
            & 1 \\[2mm]
        x^p
            & [-1,1]
            & p
            & 1 \\[2mm]
        x^{-1}
            & [1/\kappa,1]
            & \kappa^2
            & \kappa
    \end{array}
    \]
    These bounds follow directly from the derivatives of the four functions. Moreover, the corresponding families are uniformly logspace computable.

    Given that $p,\beta,\kappa=\poly(n)$, both the Lipschitz constants and the suprema in the table are polynomially bounded. To obtain any desired additive accuracy $\delta=1/\poly(n)$, apply
    \cref{lem: CM17 Lipschitz} with
    \[
        \varepsilon
            :=
            \frac{\delta}{K+M}.
    \]
    Since $K+M=\poly(n)$, this choice still satisfies $\varepsilon=\Omega(1/\poly(n))$, and the resulting additive error is precisely $\delta$.
\end{proof}

\subsection{DQC1-hardness}

We now prove matching hardness for the normalized trace of a polynomial power and for the normalized trace of the inverse. Both reductions use the same circuit-to-Hamiltonian construction of Brand{\~a}o \cite{Bra08}. The construction encodes the rejection probability of a $\DQC1$ circuit in the average of a low-energy spectral band and separates that band from the rest of the spectrum by a controllable gap. We then apply a spectral transformation that is approximately affine on the low-energy band and suppresses the contribution of the remaining eigenvalues.

\subsubsection{The circuit-to-Hamiltonian construction}

Let
\[
    Q=U_T\cdots U_1
\]
be a $\DQC1$ circuit acting on $n$ qubits and consisting of $T$ gates. By padding with identity gates, we may assume that $T+1$ is a power of two. 
Brand{\~a}o \cite{Bra08} considered the following Hamiltonian, inspired by the original construction of Kitaev \cite{KSV02}, acting on~\mbox{$N=n+\log(T+1)$} qubits:
\[
    H_{\DQC1}
        :=H_{\mathrm{out}}
            +J_{\mathrm{in}}H_{\mathrm{in}}
            +J_{\mathrm{prop}}H_{\mathrm{prop}},
\]
where
\begin{align*}
    H_{\mathrm{out}}
        &:= (T+1)\ket{0}\bra{0}_1\otimes I\otimes
            \ket{T}\bra{T}_{\mathrm{clock}},\\
    H_{\mathrm{in}}
        &:= \ket{1}\bra{1}_1\otimes I\otimes
            \ket{0}\bra{0}_{\mathrm{clock}},\\
    H_{\mathrm{prop}}
        &:=\sum_{t=1}^T H_{\mathrm{prop},t},\\
    H_{\mathrm{prop},t}
        &:= I\otimes\ket{t-1}\bra{t-1}
            +I\otimes\ket{t}\bra{t}
            -U_t\otimes\ket{t}\bra{t-1}
            -U_t^\dagger\otimes\ket{t-1}\bra{t}.
\end{align*}
The binary-clock representation makes $H_{\DQC1}$ log-local. The property of the construction used below is the following.

\begin{theorem}[\cite{Bra08}]
\label{thm:H_DQC1}
    There is a universal constant $\delta>0$ such that the following holds.
    For every $T,\Delta,\eta^{-1}\geq1$, there are coupling strengths
    \[
        J_{\mathrm{in}},J_{\mathrm{prop}}
            = \poly(T,\Delta,1/\eta)
    \]
    such that
    \[
        \left|\frac{1}{2^{n-1}}\sum_{i=1}^{2^{n-1}}\lambda_i(H_{\DQC1})-\mu_{\mathrm{reject}}(Q)\right|
            \leq\eta,
    \]
    and
    \[
        \lambda_{2^{n-1}}(H_{\DQC1})\leq\delta,
            \qquad \lambda_{2^{n-1}+1}(H_{\DQC1})\geq\Delta.
    \]
\end{theorem}

Thus, the construction produces a low-energy band of exactly $2^{n-1}$ eigenvalues whose average approximates $\mu_{\mathrm{reject}}(Q)$, while every remaining eigenvalue lies above the freely chosen threshold~$\Delta$.

For any scalar function $f$, the normalized spectral sum can be split according to these two bands as
\begin{align*}
    2(T+1)\frac{1}{2^N}\tr[f(H_{\DQC1})]
        &=\frac{1}{2^{n-1}}
            \sum_{i=1}^{2^{n-1}}f(\lambda_i(H_{\DQC1}))
                \;+\;\frac{1}{2^{n-1}}
                \sum_{i=2^{n-1}+1}^{2^N}f(\lambda_i(H_{\DQC1})).
\end{align*}
The first term reveals the average low-lying eigenvalue when $f$ is nearly affine on $[0,\delta]$, the second can be made negligible when $f$ sufficiently suppresses eigenvalues above $\Delta$. Brand{\~a}o applied this principle with $f(x)=e^{-\beta x}$ to prove $\DQC1$-hardness of normalized partition-function estimation. We use the two transformations illustrated in \cref{fig:power-inverse-spectral-transformations} to prove $\DQC1$-hardness of estimating traces of powers and inverses.

\begin{figure}[H]
    \centering

    \begin{minipage}[t]{0.48\textwidth}
        \centering
        \begin{tikzpicture}
            \begin{axis}[
                width=\linewidth,
                height=5.7cm,
                xmin=0,
                xmax=1.03,
                ymin=0,
                ymax=1.08,
                axis lines=left,
                axis line style={->},
                xlabel={transformed eigenvalues},
                ylabel={},
                xtick=\empty,
                ytick=\empty,
                clip=false,
                title={
                    $\widehat H_{\mathrm{pow}}
                        =I-\beta H_{\DQC1}$,
                    \quad $f(x)=x^p$
                },
                title style={
                    font=\small,
                    align=center,
                    yshift=2pt
                },
                label style={font=\small},
            ]

                \path[
                    fill=red!14,
                    draw=red!55!black,
                    dashed
                ]
                    (axis cs:0.14,0)
                    rectangle
                    (axis cs:0.55,1.04);

                \path[
                    fill=forestgreen!20,
                    draw=forestgreen!70!black,
                    dashed
                ]
                    (axis cs:0.78,0)
                    rectangle
                    (axis cs:0.96,1.04);

                \addplot[
                    blue,
                    very thick,
                    domain=0.04:1,
                    samples=120
                ]
                    {x^5};

                \node[
                    red!65!black,
                    align=center,
                    font=\scriptsize
                ]
                    at (axis cs:0.345,0.18)
                    {remaining\\spectrum};

                \node[
                    forestgreen!60!black,
                    align=center,
                    font=\scriptsize
                ]
                    at (axis cs:0.87,0.18)
                    {encoded\\$\mu_{\mathrm{reject}}$};

                \node[
                    blue,
                    font=\scriptsize,
                    anchor=west
                ]
                    at (axis cs:0.60,0.22)
                    {$x^p$};

            \end{axis}
        \end{tikzpicture}

        \vspace{-0.1cm}
        \textbf{(a) Trace of a power}
    \end{minipage}
    \hfill
    \begin{minipage}[t]{0.48\textwidth}
        \centering
        \begin{tikzpicture}
            \begin{axis}[
                width=\linewidth,
                height=5.7cm,
                xmin=0.14,
                xmax=1.02,
                ymin=0,
                ymax=5.55,
                axis lines=left,
                axis line style={->},
                xlabel={transformed eigenvalues},
                ylabel={},
                xtick=\empty,
                ytick=\empty,
                clip=false,
                title={
                    $\widehat H_{\mathrm{inv}}
                        =\kappa^{-1}I+\beta H_{\DQC1}$,
                    \quad $f(x)=x^{-1}$
                },
                title style={
                    font=\small,
                    align=center,
                    yshift=2pt
                },
                label style={font=\small},
            ]

                \path[
                    fill=forestgreen!20,
                    draw=forestgreen!70!black,
                    dashed
                ]
                    (axis cs:0.18,0)
                    rectangle
                    (axis cs:0.32,5.35);

                \path[
                    fill=red!14,
                    draw=red!55!black,
                    dashed
                ]
                    (axis cs:0.55,0)
                    rectangle
                    (axis cs:0.91,5.35);

                \addplot[
                    blue,
                    very thick,
                    domain=0.18:1,
                    samples=150
                ]
                    {1/x-0.1};

                \node[
                    forestgreen!60!black,
                    align=center,
                    font=\scriptsize
                ]
                    at (axis cs:0.25,0.75)
                    {encoded\\$\mu_{\mathrm{reject}}$};

                \node[
                    red!65!black,
                    align=center,
                    font=\scriptsize
                ]
                    at (axis cs:0.73,0.75)
                    {remaining\\spectrum};

                \node[
                    blue,
                    font=\scriptsize,
                    anchor=west
                ]
                    at (axis cs:0.4,2.65)
                    {$x^{-1}$};

            \end{axis}
        \end{tikzpicture}

        \vspace{-0.1cm}
        \textbf{(b) Trace of the inverse}
    \end{minipage}

    \caption{
        Schematic comparison of the two spectral transformations used in our hardness reductions. In both reductions, the eigenvalues encoding the rejection probability are mapped close to a region where the relevant function is well approximated by an affine function, while the remaining spectrum is mapped to a region in which its contribution is suppressed.
    }
    \label{fig:power-inverse-spectral-transformations}
\end{figure}

\subsubsection{Hardness of traces of powers}

We first implement the template with the power function.

\begin{theorem}
\label{thrm: trace H^p is DQC1-hard}
    The following promise problem is $\DQC1$-hard:\\
    Given a log-local Hamiltonian~\mbox{$H\in\mathbb{C}^{2^n\times2^n}$}, an integer $p$, and threshold values $a<b$ satisfying
    \[
        \sigma(H)\subseteq[0,1],
            \qquad p=\poly(n),
            \qquad b-a\geq\frac{1}{\poly(n)},
    \]
    decide whether
    \[
        \frac{1}{2^n}\tr[H^p]\geq b
            \qquad\text{or}\qquad
                \frac{1}{2^n}\tr[H^p]\leq a.
    \]
\end{theorem}

\begin{proof}
    Let $Q=U_T\cdots U_1$ be a $\DQC1$ circuit acting on $n$ qubits with $T=\poly(n)$ gates. We reduce from the standard circuit-value promise problem of deciding whether
    \[
        \mu_{\mathrm{reject}}(Q)\geq b_Q
            \qquad\text{or}\qquad
                \mu_{\mathrm{reject}}(Q)\leq a_Q,
    \]
    where $b_Q-a_Q\geq1/\poly(n)$, and pad with identity gates to assume that $T+1$ is a power of two.

    \emph{Choice of parameters.}
    Let $\delta>1$ be the universal constant from \cref{thm:H_DQC1}, and set
    \[
        \eta:=\frac{b_Q-a_Q}{3(2+\delta^2/2)}.
    \]
    Then $\eta=1/\poly(n)$. Choose
    \[
        \Delta
            :=\left\lceil
                \frac{1}{\eta}\log\!\left(\frac{2(T+1)}{\eta^2}\right)
                \right\rceil
            \leq \poly(n),      
    \]
    and apply \cref{thm:H_DQC1} with accuracy $\eta$ to get coupling strengths $J_{\mathrm{in}},J_{\mathrm{prop}}=\poly(n)$ such that
    \begin{equation}
    \label{eq:power-low-average}
        \left|\frac{1}{2^{n-1}}\sum_{i=1}^{2^{n-1}}\lambda_i-\mu_{\mathrm{reject}}(Q)\right|
            \leq\eta,
            \qquad \text{and}\qquad
        \lambda_{2^{n-1}}
            \leq\delta
            <\Delta
            \leq\lambda_{2^{n-1}+1},
    \end{equation}
    where $\lambda_i = \lambda_i(H_{\DQC1})$.

    Define
    \[
        p:=\left\lceil
            (T+1)+J_{\mathrm{in}}+4J_{\mathrm{prop}}
        \right\rceil,
        \qquad
        \beta:=\frac{\eta}{p},
        \qquad
        \widehat H:=I-\beta H_{\DQC1}.
    \]
    We find that $0\leq\beta\lambda_i \leq \eta<1$, and therefore
    \[
        \sigma(\widehat H)\subseteq[1-\eta,1]\subseteq[0,1].
    \]
    Moreover, $\widehat H$ is log-local, acts on $N=n+\log(T+1)$ qubits, and $p=\poly(N)$.

    \emph{Relating the spectral sum to the rejection probability.}
    Write
    \[
        \tau_{\mathrm{pow}}
            :=\frac{1}{2^N}\tr[\widehat H^p].
    \]
    Separating the lowest $2^{n-1}$ eigenvalues and using $2^N=2^n(T+1)$ gives
    \[
        \left| 2(T+1)\tau_{\mathrm{pow}} - \frac{1}{2^{n-1}}\sum_{i=1}^{2^{n-1}}(1-\beta\lambda_i)^p \right|
        \leq 2(T+1)(1-\beta\Delta)^p.
    \]
    Since $1-x\leq e^{-x}$ and $p\beta=\eta$, the choice of $\Delta$ yields
    \begin{equation}
    \label{eq:power-tail-decision}
        2(T+1)(1-\beta\Delta)^p
            \leq2(T+1)e^{-\eta\Delta}
            \leq\eta^2.
    \end{equation}
    Taylor's theorem at $1$ gives, for $x\in[0,1]$,
    \[
        \left|x^p-\bigl(1+p(x-1)\bigr)\right|
            \leq\frac{p^2}{2}(x-1)^2.
    \]
    Since $\lambda_i\leq\delta$ for $i\leq2^{n-1}$, combining this with
    \cref{eq:power-tail-decision} gives
    \[
        \left|
            2(T+1)\tau_{\mathrm{pow}}
            -\bigl(1-\eta\overline\lambda\bigr)
        \right|
        \leq \eta^2 + \frac{p^2}{2}\left((1-\beta\delta)-1\right)^2
        = \eta^2\left(1+\frac{\delta^2}{2}\right),
    \]
    where $\overline{\lambda} = \frac{1}{2^{n-1}}\sum_{i=1}^{2^{n-1}}\lambda_i$.
    Hence, for
    \[
        F_{\mathrm{pow}}(x)
            :=\frac{1-2(T+1)x}{\eta},
    \]
    we obtain from \cref{eq:power-low-average},
    \begin{equation}
    \label{eq:power-affine-recovery}
        \left|
            F_{\mathrm{pow}}(\tau_{\mathrm{pow}})
            -\mu_{\mathrm{reject}}(Q)
        \right|
        \leq\left(2+\frac{\delta^2}{2}\right)\eta
        =:r.
    \end{equation}
    By our choice of $\eta$, $r=(b_Q-a_Q)/3$.

    \emph{Target thresholds.}
    Define
    \[
        a_{\mathrm{pow}}
            :=\frac{1-\eta(b_Q-r)}{2(T+1)},
        \qquad
        b_{\mathrm{pow}}
            :=\frac{1-\eta(a_Q+r)}{2(T+1)}.
    \]
    Since $F_{\mathrm{pow}}$ is decreasing,
    \cref{eq:power-affine-recovery} yields that
    \[
        \mu_{\mathrm{reject}}(Q)\leq a_Q
            \quad\text{implies}\quad
        \tau_{\mathrm{pow}}\geq b_{\mathrm{pow}},
    \]
    whereas
    \[
        \mu_{\mathrm{reject}}(Q)\geq b_Q
            \quad\text{implies}\quad
        \tau_{\mathrm{pow}}\leq a_{\mathrm{pow}}.
    \]
    Furthermore,
    \[
        b_{\mathrm{pow}}-a_{\mathrm{pow}}
            =\frac{\eta(b_Q-a_Q-2r)}{2(T+1)}
            =\frac{\eta(b_Q-a_Q)}{6(T+1)}
            \geq\frac{1}{\poly(N)}.
    \]
    This completes the reduction.
\end{proof}

\subsubsection{Hardness of traces of inverses}

For the inverse, we shift the low-energy band close to $1/\kappa$.

\begin{theorem}
\label{thm: trace A inverse is DQC1-hard}
    The following promise problem is $\DQC1$-hard:\\
    Given a log-local Hamiltonian~$H\in\mathbb{C}^{2^n\times2^n}$, a parameter $\kappa\geq2$, and threshold values $a<b$ satisfying
    \[
        \sigma(H)\subseteq[1/\kappa,1],
            \qquad \kappa=\poly(n),
            \qquad b-a\geq\frac{1}{\poly(n)},
    \]
    decide whether
    \[
        \frac{1}{2^n}\tr[H^{-1}]\geq b
            \qquad\text{or}\qquad
                \frac{1}{2^n}\tr[H^{-1}]\leq a.
    \]
\end{theorem}

\begin{proof}
    Let $Q=U_T\cdots U_1$ be a $\DQC1$ circuit acting on $n$ qubits with
    $T=\poly(n)$ gates. As above, we reduce from deciding whether
    \[
        \mu_{\mathrm{reject}}(Q)\leq a_Q
            \qquad\text{or}\qquad
                \mu_{\mathrm{reject}}(Q)\geq b_Q,
    \]
    where $b_Q-a_Q\geq1/\poly(n)$, and pad the circuit so that $T+1$ is a
    power of two.

    \emph{Choice of parameters.}
    Let $\delta>1$ be the universal constant from \cref{thm:H_DQC1}, and set
    \[
        \eta:=\frac{b_Q-a_Q}{3(2+\delta^2)}.
    \]
    Choose
    \[
        \Delta:=\left\lceil\frac{2(T+1)}{\eta^3}\right\rceil
            \leq \poly(n),
    \]
    and apply \cref{thm:H_DQC1} with accuracy $\eta$ to obtain $J_{\mathrm{in}},J_{\mathrm{prop}}=\poly(n)$ such that
    \begin{equation}
    \label{eq:inverse-low-average}
        |\overline\lambda-\mu_{\mathrm{reject}}(Q)|\leq\eta,
            \qquad \text{and}\qquad
                \lambda_{2^{n-1}}\leq\delta
                    < \Delta
                    \leq \lambda_{2^{n-1}+1},
    \end{equation}
    where again $\lambda_i=\lambda_i(H_{\DQC1})$ and $\overline{\lambda} = \frac{1}{2^{n-1}}\sum_{i=1}^{2^{n-1}}\lambda_i$.

    Define
    \[
        \kappa:=(T+1)+J_{\mathrm{in}}+4J_{\mathrm{prop}},
        \qquad
        \beta:=\frac{\eta}{\kappa},
        \qquad
        \widehat H:=\frac{1}{\kappa}I+\beta H_{\DQC1}.
    \]
    Since $\|H_{\DQC1}\|_2\leq\kappa$, every eigenvalue of $\widehat H$ is at least $1/\kappa$, and upper bounded by $1/\kappa+\eta$.
    As such, we indeed find that
    \[
        \sigma(\widehat H) \subseteq [1/\kappa,1].
    \]
    Furthermore, the Hamiltonian $\widehat H$ is log-local, acts on $N=n+\log(T+1)$ qubits, and $\kappa=\poly(N)$.

    \emph{Relating the spectral sum to the rejection probability.}
    Write
    \[
        \tau_{\mathrm{inv}}
            :=\frac{1}{2^N}\tr[\widehat H^{-1}],
        \qquad
        \widehat\lambda_i
            :=\frac{1}{\kappa}+\beta\lambda_i.
    \]
    Separating the two spectral bands gives
    \[
        \left|2(T+1)\tau_{\mathrm{inv}}-\frac{1}{2^{n-1}}\sum_{i=1}^{2^{n-1}}\frac{1}{\widehat\lambda_i}\right|
        \leq\frac{2(T+1)}{\beta\Delta}.
    \]
    Taylor's theorem at $1/\kappa$ gives, for $x\in[1/\kappa,1]$,
    \[
        \left|
            \frac{1}{x}
            -\left(\kappa-\kappa^2
                \left(x-\frac{1}{\kappa}\right)\right)
        \right|
        \leq\kappa^3
            \left(x-\frac{1}{\kappa}\right)^2.
    \]
    Since the low-energy eigenvalues satisfy $\lambda_i\leq\delta$, we obtain
    \[
        \left|
            2(T+1)\tau_{\mathrm{inv}}
            -\bigl(\kappa-\kappa^2\beta\overline\lambda\bigr)
        \right|
        \leq
        \frac{2(T+1)}{\beta\Delta}
            +\kappa^3\beta^2\delta^2.
    \]
    Hence, for
    \[
        F_{\mathrm{inv}}(x)
            :=\frac{\kappa-2(T+1)x}{\kappa^2\beta},
    \]
    the choice $\beta=\eta/\kappa$ and the definition of $\Delta$ imply
    \[
        \left|F_{\mathrm{inv}}(\tau_{\mathrm{inv}})
            -\overline\lambda\right|
        \leq\frac{2(T+1)}{\eta^2\Delta}+\eta \delta^2
        \leq(1+\delta^2)\eta.
    \]
    Together with \cref{eq:inverse-low-average}, this gives
    \begin{equation}
    \label{eq:inverse-affine-recovery}
        \left|
            F_{\mathrm{inv}}(\tau_{\mathrm{inv}})
            -\mu_{\mathrm{reject}}(Q)
        \right|
        \leq(2+\delta^2)\eta
        =:r.
    \end{equation}
    By our choice of $\eta$, $r=(b_Q-a_Q)/3$.

    \emph{Target thresholds.}
    Define
    \[
        a_{\mathrm{inv}}
            :=\frac{\kappa-\kappa\eta(b_Q-r)}{2(T+1)},
        \qquad
        b_{\mathrm{inv}}
            :=\frac{\kappa-\kappa\eta(a_Q+r)}{2(T+1)},        
    \]
    Since $F_{\mathrm{inv}}$ is decreasing,
    \cref{eq:inverse-affine-recovery} yields that
    \[
        \mu_{\mathrm{reject}}(Q)\leq a_Q
            \quad\text{implies}\quad
        \tau_{\mathrm{inv}}\geq b_{\mathrm{inv}},
    \]
    whereas
    \[
        \mu_{\mathrm{reject}}(Q)\geq b_Q
            \quad\text{implies}\quad
        \tau_{\mathrm{inv}}\leq a_{\mathrm{inv}}.
    \]
    Finally,
    \[
        b_{\mathrm{inv}}-a_{\mathrm{inv}}
            =\frac{\kappa\eta(b_Q-a_Q-2r)}{2(T+1)}
            =\frac{\kappa\eta(b_Q-a_Q)}{6(T+1)}
            \geq\frac{1}{\poly(N)}.\qedhere
    \]
\end{proof}

\section{PP-completeness of estimating unnormalized spectral sums}\label{sec: PP-completeness}

We now turn from normalized spectral sums to unnormalized spectral sums. In that setting, the sampling algorithms from \cref{sec:classical_algorithm} no longer run efficiently. Nevertheless, the spectral sums can be represented as the sign of a $\GapP$ function, leading to a $\PP$ upper bound, even for inverse exponential accuracy.

Throughout this section, all numerical input data, including matrix entries, polynomial coefficients, and thresholds, are assumed to be rational numbers of polynomial bit complexity.

\subsection{PP-containment}

We begin with the case where $f=p$ is a polynomial. Although the closed-walk expansion of~$\tr[p(A)]$ contains exponentially many terms, each term is specified by only polynomially many bits and its weight can be computed in polynomial time. After a sufficiently fine discretization, their signed sum can therefore be encoded by a $\GapP$ function.

\begin{lemma}
\label{lem: PP-containment}
    Let $p_n(x)$ be a family of real polynomials of degree $d\leq \poly(n)$ and coefficient norm $\|p_n\|_1 \leq 2^{\poly(n)}$.
    Then, the following problem is contained in $\PP$:
    Given sparse access to an $s$-sparse Hermitian matrix~\mbox{$A\in\mathbb{C}^{2^n\times 2^n}$}, where $s\leq \poly(n)$, $\|A\|_2\leq 1$, and given threshold values $a<b$ such that
    \[
        \Delta:=b-a\geq2^{-\poly(n)},
    \]
    decide whether
    \[
        \tr[p_n(A)]\geq b
        \qquad\text{or}\qquad
        \tr[p_n(A)]\leq a.
    \]
\end{lemma}
\begin{proof}
    We use the standard characterization of $\PP$ by the sign of a $\GapP$ function, that is, recall, a function that maps an input string $x$ to the difference of the number of accepting and rejecting paths of a non-deterministic polynomial-time Turing machine run on $x$.
    A promise problem~\mbox{$L=(L_{\mathrm{yes}}, L_{\mathrm{no}})$} belongs to $\PP$ if there exists a function $g\in\GapP$ such that
    \begin{itemize}
        \item $x\in L_{\mathrm{yes}}$ implies $g(x)>0$, and
        \item $x\in L_{\mathrm{no}}$ implies $g(x) <0$.
    \end{itemize}

    We write \mbox{$p(x) = p_n(x) = \sum_{k=0}^d c_k x^k$} for the stated polynomial.
    Observe that for every $k$ in~$\{0,\ldots,d\}$ and $i\in[2^n]$, we have the closed-walk expansion
    \[
        A^k(i,i) 
            = \sum_{w\in W_i^{(k)}} \prod_{\ell=0}^{k-1} A(w_\ell,w_{\ell+1}),
    \]
    where $W_i^{(k)}$ denotes the set of length-$k$ walks beginning and ending at $i$ whose consecutive edges correspond to non-zero entries in $A$.
    Consequently,
    \begin{equation}\label{eq:trace-closed-walk-expansion}
        \tr[p(A)]
            = \sum_{k=0}^d \sum_{i\in[2^n]} \sum_{w\in W_i^{(k)}}
                \Re\left[c_k\prod_{\ell=0}^{k-1}A(w_\ell,w_{\ell+1})\right].
    \end{equation}
    The real part may be inserted because $A$ is Hermitian and $p$ has real coefficients, and hence~$\tr[p(A)]$ is real.

    A polynomial-length bit-string $\omega\in\{0,1\}^q$ can be uniquely associated to every summand in~\cref{eq:trace-closed-walk-expansion} by encoding $k\in\{0,\ldots,d\}$, $i\in[2^n]$ and up to $d$ local neighbor indices $j_\ell \in [s]$.
    For such a bit-string $\omega \in \{0,1\}^q$, let $r_\omega$ be the real weight of the corresponding valid summand.
    In case~$\omega\in\{0,1\}^q$ is an invalid encoding that is not associated to any summand, we let $r_\omega := 0$.
    Then
    \begin{equation}\label{eq:trace-summands}
        \tr[p(A)]
            = \sum_{\omega\in\{0,1\}^q}r_\omega.
    \end{equation}

    Since $\|A\|_2\leq1$, every matrix entry satisfies $|A(i,j)|\leq1$, and therefore
    \[
        |r_\omega|\leq \|p\|_1
            \qquad \text{for every }\omega\in\{0,1\}^q.
    \]
    Further, choose a power of two $Q$ satisfying
    \begin{equation}\label{eq:choice-Q}
        Q > \frac{\|p\|_1(2^q+1/2)}{\Delta}.
    \end{equation}
    Since $\|p\|_1,2^q,\Delta^{-1}\leq2^{\poly(n)}$, the binary representation of $Q$ has polynomial length.
    
    For every bit-string $\omega$, we define
    \[
        R_\omega := \left\lfloor \frac{Q r_\omega}{\|p\|_1} \right\rceil.
    \]
    The number $R_\omega$ can be computed in polynomial time, and satisfies
    \[
        -Q\leq R_\omega\leq Q,
            \qquad \left|R_\omega-\frac{Q r_\omega}{\|p\|_1}\right|
                \leq\frac12.
    \]

    Now, consider the following nondeterministic polynomial-time machine. 
    It first guesses~\mbox{$\omega\in\{0,1\}^q$} and then guesses
    \[
        t\in\{0,\ldots,2Q-1\}.
    \]
    It accepts if
    \[
        t<Q+R_\omega
    \]
    and rejects otherwise. 
    For a fixed $\omega$, the machine therefore has $Q+R_\omega$ accepting paths and~$Q-R_\omega$ rejecting paths. 
    The total acceptance gap of the machine is consequently
    \[
        G := 2\sum_{\omega\in\{0,1\}^q}R_\omega,
    \]
    and hence $G\in\GapP$. 
    By \cref{eq:trace-summands},
    \begin{align}
    \label{eq:gap-trace-approximation}
        \left|G-\frac{2Q}{\|p\|_1}\tr[p(A)]\right|
            = 2\left|\sum_{\omega\in\{0,1\}^q}\left(R_\omega-\frac{Q r_\omega}{\|p\|_1}\right)\right|
            \leq 2^q.
    \end{align}

    Now define the midpoint of the two promise thresholds by
    \[
        m:=\frac{a+b}{2},
            \qquad \text{and let}\qquad
                B := \left\lfloor \frac{2Qm}{\|p\|_1} \right\rceil.
    \]
    Clearly, $B$ is computable in polynomial time and has polynomially many bits. Since integer-valued polynomial-time computable functions belong to $\GapP$, and $\GapP$ is closed under subtraction, the function
    \[
        g:=G-B
    \]
    also belongs to $\GapP$. 
    Furthermore,
    \begin{equation}\label{eq:centered-gap}
        \left|g-\frac{2Q}{\|p\|_1}\left(\tr[p(A)]-m\right)\right|
            \leq 2^q+\frac12.
    \end{equation}

    Suppose now that $\tr[p(A)]\geq b$. Since $b-m=\Delta/2$, it follows from the choice of $Q$ in \cref{eq:choice-Q} and \cref{eq:centered-gap} that
    \begin{align*}
        g \geq \frac{Q\Delta}{\|p\|_1} - 2^q-\frac12
            > \left(2^q + \frac{1}{2}\right)-2^q-\frac12
            =0.
    \end{align*}
    Conversely, if $\tr[p(A)]\leq a$, then $a-m=-\Delta/2$, and therefore
    \begin{align*}
        g \leq -\frac{Q\Delta}{\|p\|_1} + 2^q+\frac12
            < -\left(2^q+\frac{1}{2}\right)+2^q+\frac12
            =0.
    \end{align*}

    Thus, the sign of the $\GapP$ function $g$ distinguishes the two promised cases. By the $\GapP$ characterization of $\PP$, the problem belongs to $\PP$.
\end{proof}

We next pass from polynomials to general spectral functions. The only additional point is the required approximation scale. Since the trace is unnormalized, a uniform approximation error $\varepsilon$ can accumulate over all $2^n$ eigenvalues and produce trace error as large as $2^n\varepsilon$. We therefore approximate $f_n$ to accuracy proportional to $2^{-n}$ times the promise gap. The assumptions below ensure that the resulting polynomial still has polynomial degree and at most exponential coefficient norm, so that the preceding lemma applies.

\begin{theorem}
\label{thm: PP-containment}
    Let $I_n\subseteq[-1,1]$ and let $f_n:I_n\to\mathbb{R}$ be a family of functions.
    Suppose that for every $\varepsilon\in(0,1)$ and $n\geq 1$, one can compute in time $\poly(n,\log(1/\varepsilon))$ a real polynomial~\mbox{$p(x) = \sum_{k=0}^{d} c_k x^k$} so that
    \vspace{-.1cm}
    \[
        \sup_{x\in I_n}\left| f_n(x) - p(x) \right| \leq \varepsilon,
    \]
    and which is of bounded degree and has bounded coefficient norm,\vspace{-.15cm}
    \[
        d\leq \poly\!\left(n,\log(1/\varepsilon)\right),
            \qquad
            \|p\|_1 = \sum_{k=0}^d |c_k|\leq2^{\poly\left(n,\log(1/\varepsilon)\right)}.
            \vspace{-.1cm}
    \]
    Then, the following problem is contained in $\PP$:
    Given sparse access to an $s$-sparse Hermitian matrix $A\in\C^{2^n\times2^n}$, where $s=\poly(n)$ and where the spectrum of $A$ is contained in $I_n$, and threshold values $a<b$ such that
    \[
        \Delta:=b-a\geq2^{-\poly(n)},
    \]
    decide whether
    \[
        \tr[f_n(A)]\geq b
        \qquad\text{or}\qquad
        \tr[f_n(A)]\leq a.
    \]
\end{theorem}
\begin{proof}
    Set
    \[
        \varepsilon:=\frac{\Delta}{3\cdot2^n}.
    \]
    Since $\Delta\geq2^{-\poly(n)}$, we have $\log(1/\varepsilon)\leq\poly(n)$. 
    By assumption, we can compute a polynomial~$p(x)$ of degree~$\poly(n)$ and bounded coefficient norm $\|p\|_1\leq 2^{\poly(n)}$ such that
    \[
        \sup_{x\in I_n}|f_n(x)-p(x)|\leq\varepsilon.
    \]
    A simple calculation shows
    \begin{equation}
    \label{eq:trace-functional-approximation}
        \bigl|\tr[f_n(A)]-\tr[p(A)]\bigr|
            \leq2^n\sup_{x\in I_n}|f_n(x)-p(x)|
            \leq\frac{\Delta}{3}.
    \end{equation}
    Consequently,
    \[
        \tr[f_n(A)]\geq b
            \qquad\text{implies}\qquad
        \tr[p(A)]\geq b-\frac{\Delta}{3},
    \]
    whereas
    \[
        \tr[f_n(A)]\leq a
            \qquad\text{implies}\qquad
        \tr[p(A)]\leq a+\frac{\Delta}{3}.
    \]
    The new thresholds are separated by $\Delta/3\geq2^{-\poly(n)}$.
    Containment in $\PP$ therefore follows directly from \cref{lem: PP-containment}.
\end{proof}

\subsection{PP-hardness}

We now prove our $\PP$-hardness result by a direct reduction from $\mathrm{MAJSAT}$.

\begin{theorem}
\label{thm: PP-hardness}
Let $I_n\subseteq[-1,1]$ and let $f_n:I_n\to\mathbb{R}$ be a family of polynomial-time computable functions.
Suppose there exist polynomial-time computable points $\alpha_n,\beta_n\in I_n$ and a gap~$\Delta_n>0$ so that for all $n\geq 1$,
\[
    \left| f_n(\alpha_n) - f_n(\beta_n)\right| \geq \Delta_n.
\]
Then, the following problem is $\PP$-hard:
Given sparse access to an $s$-sparse Hermitian matrix~\mbox{$A\in\C^{2^n\times2^n}$}, where $s=\poly(n)$ and where the spectrum of $A$ is contained in $I_n$, and threshold values $a<b$ satisfying 
\[
    b-a \geq \Delta_n,
\]
decide whether
\[
    \tr[f_n(A)]\geq b
    \qquad\text{or}\qquad
    \tr[f_n(A)]\leq a.
\]
\end{theorem}
\begin{proof}
We show a polynomial-time reduction from $\mathrm{MAJSAT}$. 
Fix $\alpha_n,\beta_n\in I_n$ as in the statement, and by exchanging $\alpha_n$ and $\beta_n$ if necessary, assume that
\[
    f_n(\alpha_n)-f_n(\beta_n) 
        \geq \Delta_n
        >0.
\]
Given a Boolean formula $\varphi$ on $n$ variables, define the $2^n \times 2^n$ dimensional diagonal matrix $A_{\varphi}$ indexed by $n$-bit strings $x\in\{0,1\}^{n}$ via
\[
    A_\varphi(x,x):=
        \begin{cases}
            \alpha_n,&\varphi(x)=1,\\
            \beta_n,&\varphi(x)=0.
        \end{cases}
\]
This matrix is one-sparse and we can efficiently access its entries by evaluating $\varphi$. 
If $N_{\mathrm{sat}}$ denotes the number of satisfying assignments, then
\begin{align*}
\label{eq:hardness-affine-count}
    \tr[f_n(A_\varphi)]
        &= f_n(\alpha_n) N_{\mathrm{sat}} + f_n(\beta_n) \left(2^n - N_{\mathrm{sat}}\right)\\
        &= \left(f_n(\alpha_n) - f_n(\beta_n)\right) N_{\mathrm{sat}} + f_n(\beta_n) \cdot2^n.
\end{align*}
Thus, deciding whether $N_{\mathrm{sat}} >2^{n-1}$ or $N_{\mathrm{sat}}\leq 2^{n-1}$ reduces to deciding whether~\mbox{$\tr[f_n(A_\varphi)] \geq b$} or~\mbox{$\tr[f_n(A_\varphi)] \leq a$} for
\begin{align*}
    b &= \left(f_n(\alpha_n) - f_n(\beta_n)\right)\cdot (2^{n-1}+1) + f_n(\beta_n)\cdot 2^n,\quad \text{and}\\
    a &= \left(f_n(\alpha_n) - f_n(\beta_n)\right)\cdot 2^{n-1} + f_n(\beta_n)\cdot 2^n,
\end{align*}
where $b-a = f_n(\alpha_n) - f_n(\beta_n) \geq \Delta_n$.
\end{proof}

Combining \cref{thm: PP-containment,thm: PP-hardness} yields a general $\PP$-completeness criterion: efficient polynomial approximability gives containment, while a separation between two efficiently computable function values gives hardness. 
The logarithm, exponential, monomial, and reciprocal functions considered throughout this paper satisfy both conditions under their respective spectral promises. 
More precisely, the polynomial approximations from \cref{lem: log-approx,lem: exp-approx,lem: monomial-approx,lem: inv-approx} yield $\PP$-containment for estimating the corresponding unnormalized spectral sums even for inverse exponential additive accuracy, while each of these functions admits efficiently computable function values that are separated by a constant gap~$\Delta_n=\Omega(1)$, yielding $\PP$-hardness already at constant additive accuracy.

\section*{Author Contribution}
All authors contributed to the discussion and development of the results. RE and AH drafted the manuscript, and all authors contributed to reviewing and revising it. The authors used GPT~5.6 to refine the text and to assist in generating pseudocode for the classical algorithms, tables and diagrams.


\section*{Acknowledgments}
The authors thank Uma Girish and Kunal Marwaha for suggesting the idea of the presented $\PP$-containment result.
They also thank Simon Apers, Ryu Hayakawa, Yupan Liu, and Harumichi Nishimura for insightful discussions. Part of the work was done while RE was visiting Nagoya University and Yukawa Institute for Theoretical Physics, and AH was visiting IRIF and Yukawa Institute for Theoretical Physics.

RE has received support under the program ``Investissement d'Avenir'' launched by the French Government and implemented by ANR, with the reference ``ANR‐22‐CMAS-0001, QuanTEdu-France''. AH is supported by JSPS KAKENHI grant No. 24H00071, 25K24674, 25K24465. FLG is supported by JSPS KAKENHI grant No.~24H00071, 25K24674, 25K24465, MEXT Q-LEAP grant No.~JPMXS0120319794, JST ASPIRE grant No.~JPMJAP2302 and JST CREST grant No.~JPMJCR24I4.

\bibliographystyle{alpha}
\bibliography{refs}


\appendix

\section{Polynomial approximations via Chebyshev Truncation}

Let $T_k$ denote the $k$-th Chebyshev polynomial of the first kind given via
\[
  T_k(\cos\theta)=\cos(k\theta).
\]
They satisfy the standard recurrence
\[
  T_0(x)=1,\qquad T_1(x)=x,\qquad T_{k+1}(x)=2xT_k(x)-T_{k-1}(x).
\]
Since every $x\in[-1,1]$ can be written as $x=\cos\theta$, this immediately implies
\[
    |T_k(x)| \leq 1 \qquad\text{for all }x\in[-1,1].
\]

We next observe some elementary facts about the $\ell_1$-norm of (Chebyshev) polynomials.

\begin{lemma}[coefficient-norm of Chebyshev polynomials]\label{lem: ell1-norm of chebyshev polynomials}
    For all $k\geq 0$, we have~\mbox{$\|T_k\|_1\leq \left(1+\sqrt{2}\right)^k$}. 
\end{lemma}

\begin{proof}
    Let $A_k=\left\lVert T_k\right\rVert_{1}$. 
    From the Chebyshev recurrence, we obtain
    \[
      A_{k+1}\leq 2A_k+A_{k-1}.
    \]
    The number $\rho=1+\sqrt2$ satisfies $\rho^2=2\rho+1$. 
    Since $A_0=1\leq \rho^0$ and $A_1=1\leq \rho$, induction gives $A_k\leq\rho^k$.
\end{proof}

\begin{lemma}[coefficient-norm after affine substitution]\label{lem: ell1-norm after affine substitution}
    Let $p$ be a polynomial of degree at most~$d$.
    Then, for any $a,b\in\mathbb{R}$,
    \[
        \|p(ax+b)\|_1
            \leq \| p\|_1(|a|+|b|)^d.
    \]
\end{lemma}

\begin{proof}
    Let $p(x)=\sum_{k=0}^d c_kx^k$. 
    Then
    \[
      \|p(ax+b)\|_1
          \leq \sum_{k=0}^d |c_k|\| (ax+b)^k\|_1
          \leq \sum_{k=0}^d |c_k|(|a|+|b|)^k
          \leq \| p\|_1(|a|+|b|)^d.\qedhere
    \]
\end{proof}

\subsection{Approximating the monomial}

\begin{restatable}[cf. Section 3.1 in \cite{SV14}, Approximation of $x^s$]{lemma}{monomialapprox}
\label{lem: monomial-approx}
    Let $s\in\mathbb{N}$, $\varepsilon\in(0,1)$ and set
    \[
      d=\left\lceil\sqrt{2s\log(2/\varepsilon)}\right\rceil.
    \]
    There is a polynomial-time computable polynomial $p_{s,d}$ of degree at most $d$ such that
    \[
      \sup_{x\in[-1,1]} \left|x^s-p_{s,d}(x)\right|\leq \varepsilon.
    \]
    Moreover, its $\ell_1$-norm is bounded by
    \[
      \|p_{s,d}\|_1
        \leq \left(1+\sqrt{2}\right)^d
        = \exp\!\left(O\!\left(\sqrt{s\log(1/\varepsilon)}\right)\right).
    \]
\end{restatable}

The explicit construction of the low-degree approximating polynomial is from Section 3.1 in~\cite{SV14}.
Our only contribution is to make the coefficient norm bound explicit.

\begin{proof}
    It is shown in \cite{SV14} that the truncated Chebyshev expansion $p_{s,d}(x) :=\sum_{k=0}^d c_k T_k(x)$, where
    \[
        c_k =
            \begin{cases}
                2^{-s}\binom{s}{s/2}, & k=0 \text{ and } s \text{ even},\\[1mm]
                2\cdot2^{-s}\binom{s}{(s+k)/2}, & k>0,\ k\equiv s \pmod 2,\\[1mm]
                0, & \text{otherwise},
            \end{cases}
    \]
    satisfies
    \[
        \sup_{x\in[-1,1]}\left|x^s - p_{s,d}(x)\right| \leq \varepsilon
    \]
    for $d= \left\lceil\sqrt{2s\log(2/\varepsilon)}\right\rceil$.

    These Chebyshev coefficients can easily be converted to monomial coefficients by using the standard recurrence of Chebyshev polynomials.
    In order to bound the coefficient-norm of $p_{s,d}$, we first observe that $\sum_{k=0}^d c_k \leq 1$ as well as $\|T_k\|_1\leq \left(1+\sqrt{2}\right)^k$ from \cref{lem: ell1-norm of chebyshev polynomials} to obtain
    \[
        \|p_{s,d}\|_1
            \leq \sum_{k=0}^d |c_k|\|T_k\|_1
            \leq \left(1+\sqrt{2}\right)^d
            = \exp\left(O\left(\sqrt{s\log(1/\varepsilon)}\right)\right).\qedhere
    \]
\end{proof}

\subsection{Approximating the exponential}

\begin{restatable}[cf. Section 3.2 in \cite{SV14}, Approximation of $e^{-\beta x}$]{lemma}{expapprox}
\label{lem: exp-approx}
Let $\beta>0$, $\varepsilon\in(0,1)$ and set
\[
  t=\left\lceil\max\{\beta e^2/2,\; \log(2/\varepsilon)\}\right\rceil,
    \qquad
        d=\left\lceil \sqrt{2 t\log(4/\varepsilon)}\right\rceil.
\]
There is a polynomial-time computable polynomial $p_{\exp}$ of degree at most $d$ such that
\[
  \sup_{x\in[0,1]}\left|e^{-\beta x}-p_{\exp}(x)\right|\leq\varepsilon.
\]
Moreover, its $\ell_1$-norm is bounded by
\[
  \|p_{\exp}\|_1
    \leq (3+3\sqrt{2})^d
    = \exp\!\left(O\!\left(\sqrt{\beta\log(1/\varepsilon)}+\log(1/\varepsilon)\right)\right).
\]
\end{restatable}

\begin{proof}
    Let us denote by $p_{k,d}(x)$ the polynomial approximation of $x^k$ of degree at most $d$ from \cref{lem: monomial-approx}.
    Following the construction from Section~3.2 of~\cite{SV14}, define
    \[
        q_{t,d}(x) := \sum_{k=0}^t e^{-\beta/2}\frac{(-\beta/2)^k}{k!}p_{k,d}(x).
    \]
    The analysis in~\cite{SV14} shows that, for the choices of $t$ and $d$ as in the statement,
    \[
        \sup_{y\in[-1,1]}\left|e^{-\beta/2-\beta/2\cdot y} - q_{t,d}(y)\right|
            = \sup_{x\in[0,1]}\left|e^{-\beta x} - q_{t,d}(2x-1)\right| 
            \leq \varepsilon.
    \]
    As such, we can easily compute the monomial coefficients of $q_{t,d}(2x-1)$ to approximate $e^{-\beta x}$ on $[0,1]$.
    It remains to bound the $\ell_1$-norm of this polynomial.
    For this, we calculate:
    \begin{align*}
        \|q_{t,d}(2x-1)\|_1
            &\leq e^{-\beta/2} \sum_{k=0}^t \left|\frac{(-\beta/2)^k}{k!}\right|  \|p_{k,d}(2x-1) \|_1 \\
            &\leq e^{-\beta/2} \underbrace{\sum_{k=0}^t \left|\frac{(-\beta/2)^k}{k!}\right|}_{\leq e^{\beta/2}} \cdot  (3+3\sqrt{2})^d\\
            &\leq (3+3\sqrt{2})^d
            = \exp\!\left(O\left(\sqrt{\beta\log(1/\varepsilon)} + \log(1/\varepsilon)\right)\right),
    \end{align*}
    where the second inequality uses that $\|p_{k,d}(x)\|_1\leq \left(1+\sqrt{2}\right)^d$ from \cref{lem: monomial-approx} together with \cref{lem: ell1-norm after affine substitution}.
\end{proof}

\subsection{Approximating the logarithm}

Before proving the polynomial approximation of the logarithm, we introduce some more tools to deal with the non-standard approximation interval of $[\frac{1}{\kappa},1]$.
The same tools will be used for the approximation of the reciprocal on $[\frac{1}{\kappa},1]$.
Throughout, let $\kappa\geq 2$.
We define variables
\[
    q := \frac{\sqrt{\kappa} - 1}{\sqrt{\kappa}+1}
        \qquad\text{and} \qquad
        c := \frac{\left(\sqrt{\kappa}+1\right)^2}{4\kappa}.
\]
Further, consider the affine map $\phi:[-1,1]\rightarrow [\frac{1}{\kappa},1]$ given via
\[
    \phi(y) := c(1+2qy+q^2).
\]
Indeed, this maps the interval $[-1,1]$ onto $[\frac{1}{\kappa},1]$ since
\[
    \phi(-1) = c(1-q)^2
        = \frac{1}{\kappa}
        \qquad\text{and}\qquad
        \phi(1) = c(1+q)^2 = 1.
\]
The inverse of $\phi$ is the affine map $\psi:[\frac{1}{\kappa},1]\rightarrow[-1,1]$ given via
\[
    \psi(x) := \frac{2x - (1+1/\kappa)}{1-1/\kappa}.
\]

\begin{observation}\label{obs: coeff norm bound of T_k(ax+b)}
    For all $k\geq 0$, we have $\|T_k(\psi(x))\|_1 \leq (7+7\sqrt{2})^k$.
\end{observation}
\begin{proof}
    Set $a=\frac{2}{1-1/\kappa}$ and $b=-\frac{1+1/\kappa}{1-1/\kappa}$ such that $\psi(x) = ax+b$.
    Note that for $\kappa\geq 2$ we have
    \[
        |a| + |b| \leq \frac{3\kappa + 1}{\kappa - 1} \leq 7.
    \]
    As such, we directly obtain from \cref{lem: ell1-norm after affine substitution} and \cref{lem: ell1-norm of chebyshev polynomials} that
    \[
        \|T_k(\psi(x))\|_1
            \leq \|T_k(x)\|_1 (|a|+|b|)^k
            \leq (1+\sqrt{2})^k \cdot 7^k
            =(7+7\sqrt{2})^k.\qedhere
    \]
\end{proof}

\begin{restatable}[Approximation of $\log x$]{lemma}{logapprox}
\label{lem: log-approx}
Let $\kappa\geq2$, $\varepsilon\in(0,1/2)$ and set
\[
    d=\left\lceil \frac{\sqrt\kappa}{2}\log\frac{2}{\varepsilon}\right\rceil.
\]
There is a polynomial-time computable polynomial $p_{\log}$ of degree at most $d$ such that
\[
  \sup_{x\in[1/\kappa,1]}\left|\log x-p_{\log}(x)\right|\leq\varepsilon.
\]
Moreover, its $\ell_1$-norm is bounded by
\[
    \|p_{\log}\|_1
        \leq \log 4+2d(7+7\sqrt{2})^{d}
        = \exp\!\left(O\!\left(\sqrt\kappa\log\frac{1}{\varepsilon}\right)\right).
\]
\end{restatable}

\begin{proof}
    For $y=\cos\theta\in[-1,1]$, we find that
    \[
        \phi(y) 
            = c(1+2qy+q^2) 
            = c(1+qe^{i\theta})(1+qe^{-i\theta}).
    \]
    Since \(q<1\), the Taylor expansion
    \[
        \log(1+z) 
            = \sum_{k=1}^{\infty}(-1)^{k+1}\frac{z^k}{k},
    \]
    for $|z|<1$, may be applied with \(z=qe^{i\theta}\).
    Hence
    \begin{align*}
        \log\phi(y)
            &= \log c + \log(1+qe^{i\theta}) + \log(1+qe^{-i\theta}) \\
            &= \log c + 2\sum_{k=1}^{\infty}(-1)^{k+1} \frac{q^k}{k}\cos(k\theta) \\
            &= \log c + 2\sum_{k=1}^{\infty}(-1)^{k+1} \frac{q^k}{k}T_k(y).
    \end{align*}
    Thus, we find that the degree~$d$ truncation 
    \[
        \widetilde{p}_{\log}(y) := \log c + 2\sum_{k=1}^{d}(-1)^{k+1}\frac{q^k}{k}T_k(y)
    \]
    gives a small truncation error
    \begin{align*}
        \sup_{y\in[-1,1]}\left| \log \phi(y) - \widetilde{p}_{\log}(y)  \right|
            &\leq 2\sum_{k=d+1}^{\infty}\frac{q^k}{k}\\
            &\leq 2 \int_{d}^{\infty} \frac{e^{-k\log(1/q)}}{k}dk\\
            &\leq 2\frac{e^{-d\log(1/q)}}{\log(1/q)d},
    \end{align*}
    where the last inequality holds since $\frac{1}{k}\leq \frac{1}{d}$ for $k\in[d,\infty)$.
    Hence, by choosing 
    \[
        d=\left\lceil \frac{1}{\log(1/q)} \log\frac{2}{\varepsilon}\right\rceil \leq \left\lceil \frac{\sqrt{\kappa}}{2} \log\frac{2}{\varepsilon}\right\rceil
    \]
    the error becomes less or equal to $\varepsilon$.
    
    We finally set\vspace{-.4cm}
    \[
        p_{\log}(x)
            := \widetilde{p}_{\log}(\psi(x))
            = \log c + 2\sum_{k=1}^{d}(-1)^{k+1}\frac{q^k}{k}T_k(\psi(x)).
    \]
    This polynomial has the same degree as $\widetilde{p}_{\log}(x)$ and $\varepsilon$-approximates $\log x$ on $[1/\kappa,1]$ since
    \begin{align*}
        \sup_{x\in[1/\kappa,1]}\left|\log(x)-p_{\log}(x)\right|
            &= \sup_{y\in[-1,1]}\left| \log\phi(y)
                - \widetilde{p}_{\log}(y) \right|
        \leq \varepsilon.
    \end{align*}
    
    It remains to bound the coefficient norm of $p_{\log}(x)$.
    Since $c=\frac{(\sqrt{\kappa}+1)^2}{4\kappa}\in[1/4,1]$ for $\kappa\geq 2$, we have
    \[
        |\log c|\leq\log 4,
            \qquad\text{as well as}\qquad
            \|T_k(\psi(x))\|_1\leq(7+7\sqrt{2})^k,
    \]
    by \cref{obs: coeff norm bound of T_k(ax+b)}.
    Hence\vspace{-.4cm}
    \begin{align*}
        \|p_{\log}\|_1
            &\leq |\log c| + 2\sum_{k=1}^{d} \frac{q^k}{k}\|T_k(\psi(x))\|_1 \\
            &\leq \log4 + 2\sum_{k=1}^{d} \frac{q^k}{k}(7+7\sqrt{2})^k \\
            &\leq \log4 + 2d(7+7\sqrt{2})^{d}.
    \end{align*}
    
    The monomial coefficients of $p_{\log}(x)$ can easily be obtained from the Chebyshev coefficients of $\widetilde{p}_{\log}(x)$ using the standard recurrence of the Chebyshev polynomials, together with the substitution \(x\mapsto\psi(x)\).
\end{proof}

\subsection{Approximating the reciprocal}

\begin{restatable}[Approximation of $1/x$]{lemma}{invapprox}
\label{lem: inv-approx}
    Let $\kappa\geq2$ and $\varepsilon\in(0,1)$ and set
    \[
      d=\left\lceil \frac{\sqrt\kappa}{2}\log\frac{\kappa}{\varepsilon}\right\rceil.
    \]
    There is a polynomial-time computable polynomial $p_{\rm rec}$ of degree at most $d$ such that
    \[
      \sup_{x\in[1/\kappa,1]}\left|\frac{1}{x}-p_{\rm rec}(x)\right|\leq\varepsilon.
    \]
    Moreover, its $\ell_1$-norm is bounded by
    \[
      \|p_{\mathrm{rec}}\|_1
        \leq \sqrt\kappa\left(1+2d(7+7\sqrt{2})^{d}\right)
        = \exp\!\left(O\!\left(\sqrt\kappa\log\frac{\kappa}{\varepsilon}\right)\right).
    \]
\end{restatable}

\begin{proof}
    Recall again that for $y=\cos\theta \in[-1,1]$ we have
    \[
        \phi(y) 
            = c(1+2qy+q^2) 
            = c(1+qe^{i\theta})(1+qe^{-i\theta}).
    \]
    Since $q<1$, the geometric series below are absolutely convergent, and we obtain
    \begin{align*}
        1 + 2\sum_{k=1}^\infty (-q)^k T_k(y) 
            &= 1 + \sum_{k=1}^\infty (-q)^k \left(e^{ik\theta} + e^{-ik\theta}\right) \\
            &= 1 + \frac{-qe^{i\theta}}{1+qe^{i\theta}} + \frac{-qe^{-i\theta}}{1+qe^{-i\theta}} \\
            &= \frac{1-q^2}{(1+qe^{i\theta})(1+qe^{-i\theta})} \\
            &= \frac{1-q^2}{1+2qy+q^2}.
    \end{align*}
    We also observe that $c(1-q^2) = \frac{1}{\sqrt{\kappa}}$.
    Hence,
    \[
        \frac{1}{\phi(y)} = \frac{1}{c(1+2qy+q^2)}
            = \sqrt{\kappa} \left( 1+2\sum_{k=1}^{\infty}(-q)^kT_k(y)\right).
    \]
    Truncating after degree $d$, we define
    \[
        \widetilde{p}_{\mathrm{rec}}(y)
            := \sqrt{\kappa} \left( 1+2\sum_{k=1}^{d}(-q)^kT_k(y) \right).
    \]
    For the truncation error, we get
    \begin{align*}
        \sup_{y\in[-1,1]}\left|\frac{1}{\phi(y)}-\widetilde{p}_{\mathrm{rec}}(y)\right|
            &\leq 2\sqrt{\kappa}\sum_{k=d+1}^{\infty}q^k \\
            &\leq 2\sqrt{\kappa} \int_{d}^{\infty} e^{-k\log(1/q)} dk \\
            &\leq \frac{2\sqrt{\kappa}}{\log(1/q)} e^{-d\log(1/q)}.
    \end{align*}
    Choosing $d=\left\lceil \frac{1}{\log(1/q)} \log\frac{\kappa}{\varepsilon} \right\rceil \leq \left\lceil \frac{\sqrt{\kappa}}{2} \log\frac{\kappa}{\varepsilon} \right\rceil$ and using that $\log(1/q) \geq 2/\sqrt{\kappa}$, we can further bound this error by
    \[
        \frac{2\sqrt{\kappa}}{\log(1/q)} e^{-d\log(1/q)}
            \leq \frac{2\sqrt{\kappa}}{2/\sqrt{\kappa}} \cdot\frac{\varepsilon}{\kappa} = \varepsilon.
    \]
    
    We set our final approximating polynomial to be
    \[
        p_{\rm rec}(x) 
            := \widetilde{p}_{\rm rec}(\psi(x))
            = \sqrt{\kappa}\left(1+2\sum_{k=1}^{d}(-q)^kT_k(\psi(x))\right).
    \]
    Then $p_{\rm rec}(x)$ $\varepsilon$-approximates $1/x$ on $[1/\kappa,1]$ and its $\ell_1$-norm satisfies
    \begin{align*}
        \|p_{\rm rec}\|_1
            &\leq \sqrt{\kappa} \left( 1+2\sum_{k=1}^{d}q^k\|T_k(\psi(x))\|_1\right) \\
            &\leq \sqrt{\kappa} \left( 1+2\sum_{k=1}^{d}q^k(7+7\sqrt{2})^k \right) \\
            &\leq \sqrt{\kappa} \left( 1+2d(7+7\sqrt{2})^{d} \right),
    \end{align*}
    where we used \cref{obs: coeff norm bound of T_k(ax+b)} and $q<1$.

    As before, the monomial coefficients of $p_{\rm rec}(x)$ can efficiently be obtained from the Chebyshev coefficients of $\widetilde{p}_{\rm rec}(x)$ using the standard recurrence for $T_k$, together with the affine substitution~$x\mapsto\psi(x)$.
\end{proof}

\end{document}